\documentclass[USenglish,oneside,twocolumn]{article}

\usepackage[utf8]{inputenc}%(only for the pdftex engine)
\usepackage[big]{dgruyter_NEW}
\cclogo{\includegraphics{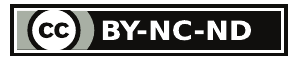}}

\hypersetup{draft}

\usepackage{etex}
\reserveinserts{28}
\usepackage{colortbl}
\usepackage{xcolor}
\usepackage{balance}
\usepackage{pifont}
\usepackage{lipsum}
\usepackage{dblfloatfix}
\usepackage{url}
\usepackage{bytefield}
\usepackage{tikz}
\usetikzlibrary{arrows,shadows,calc,positioning}
\usepackage{subcaption}
\usepackage[numbers]{natbib}
\usepackage{flushend}
\usepackage[nolist,nohyperlinks]{acronym}
\usepackage{comment}
\usepackage{xspace}
\usepackage{textcomp}
\usepackage{stackengine}
\usepackage{amsbsy}
\usepackage{amsthm}
\usepackage{amsmath}
\usepackage{algpseudocode}
\usepackage{algorithm}
\usepackage{verbatim}
\usepackage{csquotes}
\usepackage{url}
\usepackage{multirow}
\usepackage{pifont}
\usepackage{footnote}
\usepackage{amssymb}
\usepackage{booktabs}
\usepackage{adjustbox}
\usepackage{pgfmath}

\acrodefplural{OS}[OSes]{Operating Systems}
\newcommand{\xmark}{\ding{55}}%

\tikzstyle{intt}=[draw,text centered,minimum size=6em,text width=5.25cm,text height=0.34cm]
\tikzstyle{intl}=[draw,text centered,minimum size=2em,text width=2.75cm,text height=0.34cm]
\tikzstyle{int}=[draw,minimum size=2.5em,text centered,text width=3.5cm]
\tikzstyle{intg}=[draw,minimum size=3em,text centered,text width=5.cm]
\tikzstyle{sum}=[draw,shape=circle,inner sep=2pt,text centered,node distance=3.5cm]
\tikzstyle{summ}=[drawshape=circle,inner sep=4pt,text centered,node distance=3.cm]

\usepackage[justification=centering]{caption}
\newcounter{bob}

\newcommand{\etal}{\emph{et al.}\xspace}

\newcommand{\ie}{\emph{i.e.}\xspace}

\newcommand{\bi}{\begin{itemize}}
\newcommand{\ei}{\end{itemize}}
\newcommand{\be}{\begin{enumerate}}
\newcommand{\ee}{\end{enumerate}}
\newcommand{\etc}{etc.\@\xspace}

\newcommand{\lt}{layer-2\@\xspace}

\newcommand{\bco}{\begin{comment}}
\newcommand{\eco}{\end{comment}}

\begin{document}

\author*[1]{Jeremy Martin} 
\author[2]{Douglas Alpuche}         
\author[3]{Kristina Bodeman}
\author[4]{Lamont Brown}          
\author*[5]{Ellis Fenske}         
\author[6]{Lucas Foppe} 
\author*[7]{Travis Mayberry}         
\author*[8]{Erik Rye}         
\author[9]{Brandon Sipes} 
\author[10]{Sam Teplov}         

\affil[1]{The MITRE Corporation, E-mail: jbmartin@mitre.org}
\affil[2]{\ac{USNA}}
\affil[3]{\ac{USNA}}
\affil[4]{\ac{USNA}}
\affil[5]{\ac{USNA}, E-mail: fenske@usna.edu}
\affil[6]{\ac{USNA}}
\affil[7]{\ac{USNA}, E-mail: mayberry@usna.edu}
\affil[8]{CMAND, E-mail: rye@cmand.org}
\affil[9]{\ac{USNA}}
\affil[10]{\ac{USNA}}

\title{\huge Handoff All Your Privacy \textendash\ A Review of Apple's
Bluetooth Low Energy Continuity Protocol}
\runningtitle{Handoff All Your Privacy  \textendash\ A Review of Apple's
Bluetooth Low Energy Continuity Protocol}

\begin{abstract}
 {We investigate Apple's \ac{BLE} Continuity protocol,
designed to support interoperability and communication between iOS and macOS
devices, and show that the price for this seamless experience is leakage of
identifying information and behavioral data to passive adversaries. First, we
reverse engineer numerous Continuity protocol message types and identify data
fields that are transmitted unencrypted.  We show that Continuity messages are
broadcast over \ac{BLE} in response to actions such as locking and unlocking a
device's screen, copying and pasting information, making and accepting phone
calls, and tapping the screen while it is unlocked.  Laboratory experiments
reveal a significant flaw in the most recent versions of macOS that defeats
\ac{BLE} \ac{MAC} address randomization entirely by causing the public \ac{MAC}
address to be broadcast. We demonstrate that the format and content of
Continuity messages can be used to fingerprint the type and \ac{OS} version of
a device, as well as behaviorally profile users.  Finally, we show that
predictable sequence numbers in these frames can allow an adversary to track
Apple devices across space and time, defeating existing anti-tracking
techniques such as MAC address randomization. 
}
\end{abstract}


\journalname{Proceedings on Privacy Enhancing Technologies}
\DOI{Editor to enter DOI}
\startpage{1}
\received{..}
\revised{..}
\accepted{..}

\journalyear{..}
\journalvolume{..}
\journalissue{..}

\maketitle

\section{Introduction}
\label{sec:intro}
\vspace{-2mm}
The ubiquity of wirelessly connected mobile devices in the day-to-day lives of
people globally has brought with it unprecedented risk of privacy
violation for modern consumers.  Mobile devices constantly transmit and
receive information even while not in active use, and many of the
protocols driving this communication are not designed with privacy in mind.

Tracking concerns and privacy leakages in 802.11 Wi-Fi are well-known and have
been extensively studied over the last decade.  Since Wi-Fi clients must
actively probe for nearby access points to connect to, an adversary can listen
to these probes and use the device's \ac{MAC} address (which is included in
probes) to identify and track it as it moves from place to place.  This is not
an academic threat: there are multimillion-dollar companies~\cite{holger_2018,
ftc} whose business model relies on using Wi-Fi tracking data for targeted
marketing, and they control large networks of Wi-Fi access points that gather
information on all nearby devices. Users are largely unaware that these
widely-deployed tracking capabilities exist and that their Wi-Fi devices might
be leaking sensitive data. 

In response to this threat, device and \ac{OS} manufacturers began to provide
MAC address randomization as a privacy enhancement.  Rather than using the same
\ac{MAC} address consistently, which enables correlation over multiple
observations, devices employing \ac{MAC} randomization instead choose random
values, and change them periodically.  While the principle itself is sound,
many implementations of \ac{MAC} address randomization have proven ineffective
in practice~\cite{martin2017study, vanhoef2016mac}. Defeating \ac{MAC} address
randomization is largely possible due not to flaws in Wi-Fi itself, but because
of extraneous information in higher-layer protocols.  Many
technologies are not privacy-aware and leak information that can be used to
track users and devices, despite the \ac{MAC} address being effectively
hidden through randomization. 

Bluetooth, in both of its current protocol instantiations, also uses MAC
addresses as hardware identifiers.  \ac{BLE}, which we examine exclusively in
this study, has included mechanisms for a device to generate and use random
\ac{MAC} addresses, enhancing the potential privacy benefit to clients by
increasing the difficulty of tracking unique devices.  Unfortunately, it also
suffers from the same problem as Wi-Fi: manufacturers and \acp{OS} implement
features built on top of \ac{BLE} that leak sensitive information which can be
used to track users~\cite{fawaz}, defeating the purpose of \ac{MAC}
randomization itself.

In this work we investigate one such technology -- Apple's Continuity protocol.
Continuity is designed to support the seamless transfer and synchronization of
data between multiple iOS and macOS devices.  We show that in exchange for
simplifying the user experience and allowing synchronization between devices, the
messages that comprise Continuity constantly leak sensitive information; not
only identifiers that could be used to defeat randomization and track devices
wirelessly, but also behavioral information that reveals user device activity.  

\vspace{-5mm}
\subsection{Contributions}
\label{sec:contrib}
\vspace{-3mm}

\bi
  \item To the best of our knowledge, we are the first to reverse engineer
Apple's Continuity protocol. We describe the format, contents, and behavior of
Continuity messages.
  \item We identify several recognizable messages that leak behavioral data,
including when a user locks or unlocks their phone, when they are touching the
screen, when they visit certain apps or settings pages, when they receive text
messages or answer a phone call, and even when they use the copy/paste feature.
  \item We show that observing these messages allows an adversary to accurately
fingerprint the model type and OS version of a device.  We also present a novel
application of a known, hard-to-detect, active reconnaissance technique that
can further increase the precision of these fingerprints.
  \item Finally, we demonstrate how the predictable sequence numbers sent with
these messages can be exploited to defeat anti-tracking technologies like MAC
address randomization and detect the existence of a device nearby even after
not observing it for several days.
  \ei

\vspace{-11mm}
\section{Background and Related Work}
\label{sec:back}
\vspace{-2mm}

\subsection{Wireless Device Tracking and Privacy}
\vspace{-2mm}

Wireless \lt identifiers -- in particular, \ac{MAC} addresses -- have a rich and
lengthy history of being exploited for tracking mobile devices, and hence, their
owners. The broadcast nature of wireless communication, the tendency for
software designers to program devices to proactively advertise their
availability and seek out services, and the persistence and uniqueness of global
\ac{MAC} addresses combine to form a serious threat to user privacy: our
mobile devices constantly broadcasting trackable identifiers for all to hear without
our interaction. 

In order to detect nearby wireless networks, 802.11 Wi-Fi clients broadcast a
special type of wireless frame known as a \emph{probe request}. The purpose of
these frames is to solicit responses from local access points that provide
network connectivity. Contained in each frame is the client's source \ac{MAC}
address; this 48-bit value uniquely identifies the client seeking network
service to access points that provide it. Unfortunately, this also provides
adversaries a unique identifier to track users~\cite{Cunche2014,
abedi2013bluetooth, bonne2013wifipi, musa2012tracking, matte2017wi}. To address
the privacy concerns inherent in the use of the \ac{MAC} address assigned by
the device manufacturer (a \emph{globally unique} \ac{MAC} address),
manufacturers have shifted to broadcasting ephemeral, random \ac{MAC} addresses
while the device is in an un-associated state; unfortunately, \ac{MAC} address
randomization in 802.11 can often be defeated~\cite{vanhoef2016mac,
martin2017study}. 

\ac{MAC} addresses are used as \lt identifiers in Bluetooth communication as
well, and carry the same privacy and tracking risks as Wi-Fi \ac{MAC} addresses
~\cite{abedi2013bluetooth, haase2004bluetrack, versichele2012use,
liebig2012modelling}.  Similar to Wi-Fi, Bluetooth devices implement \ac{MAC}
address randomization as a mechanism to evade tracking and preserve user
privacy. In fact, randomized \ac{MAC} addresses have been part of the \ac{BLE}
standard since its introduction (known then as Bluetooth
Smart)~\cite{corespec}. As we focus exclusively on \ac{BLE} in this work, we
first discuss \ac{BLE} \ac{MAC} address structure and randomization in greater
detail.  

\vspace{-5mm}
\subsection{Bluetooth (Classic vs. Low Energy)}
\vspace{-3mm}

The term \emph{Bluetooth} is often used colloquially to refer to two distinct,
non-interoperable technologies.  The original Bluetooth standard is now
referred to as Bluetooth ``Classic'' to recognize its chronological precedence,
or Bluetooth \ac{BR}/\ac{EDR} in relation to the rate at which a device
implementing this protocol transmits data. \ac{BLE}, formerly known by the
marketing name ``Bluetooth Smart'', on the other hand, is so named due to the
lower power consumption needs of devices implementing it compared to Bluetooth
\ac{BR}/\ac{EDR}. Despite their typical use in connecting peripheral devices at
close range, both Bluetooth Classic and \ac{BLE} are capable of transmitting up
to 100m in an open area. The current \ac{BLE} version, 5.0, is rated up to
400m~\cite{nordicrange}. 

Generally speaking, Bluetooth Classic is preferred for applications in which a
constant flow of data occurs between the paired devices; for example, Bluetooth
headphones or speakers connected to a mobile phone would almost certainly
utilize Bluetooth Classic. \ac{BLE}, on the other hand, is typically used to
send short messages advertising a device's existence and some parameters
related to the device; for instance, Tile~\cite{tile} and related devices use
\ac{BLE} for proximity sensing. In this study, we are interested only in
Apple's \ac{BLE} implementation, as it lends itself to device tracking, \ac{OS}
fingerprinting, and activity profiling. 

While Bluetooth Classic and \ac{BLE} operate in the 2.4 GHz unlicensed
spectrum, each utilizes a different number and size of channels. \ac{BLE} uses
40 2 MHz channels; 37 of these are used to send and receive data, while the
remaining three are used to detect a device's presence by sending advertisement
frames at regular intervals~\cite{corespec}. With exception of our \ac{GATT}
queries in Section~\ref{sec:active}, we are concerned in this work with
messages continuously sent in the advertisement channels that are designed to
be received by nearby stations. 

\vspace{-4mm}
\subsection{\ac{BLE} Addressing and Randomization}
\vspace{-2mm}

Every Bluetooth interface is assigned a globally-unique, \emph{public} \ac{MAC}
address by the device manufacturer; Bluetooth \ac{MAC} addresses are EUI-48
identifiers and are obtained from the \ac{IEEE} Registration Authority in the
same manner as 802.11 \ac{MAC} addresses.  In order to provide users with a
measure of privacy and prevent privacy leakages~\cite{martin2016decomposition}
associated with revealing a public device address, devices are also permitted
to use \emph{random} device addresses, which are split into two categories --
\emph{static} random addresses and \emph{private} random addresses, the latter
of which is further divided into two subcategories.  

\emph{Static random addresses} are, as their name implies, random addresses
that are long-lived; the Bluetooth Core Specification mandates that these
\ac{BLE} \ac{MAC} addresses remain unchanged after
initialization~\cite{corespec}. Power cycling a device may change its static
random address, but changing a device's static address will cause any devices
that have previously connected to it to fail to automatically connect to the
previous static address. Static random addresses are identifiable by having the
two highest order bits set to 1, at least one of the remaining 46 low-order
bits set to 1, and at least one of the lower 46 bits set to 0 (\ie, two set
bits followed by 46 0s or 46 1s are not allowable static random addresses.)

\emph{Non-resolvable private random addresses} provide additional privacy
compared to using a public \ac{MAC} address, as the random address is used in
lieu of the public, globally-unique \ac{MAC} address of the device.
Non-resolvable random addresses are identifiable by the two most significant
bits being set to 0, and the remaining 46 lower order bits containing at least
one 0 and one 1 bit~\cite{corespec}. Finally, as their name implies,
non-resolvable private addresses do not aid in authenticating two devices to
each other.

\emph{Resolvable private random addresses} are the final type of random address in
Bluetooth, and are the type used by Apple to provide \ac{MAC}
address privacy; as such, this is the address type we consider in this work.
Resolvable private addresses provide the ability for devices to authenticate
each other based on the use of a 128-bit key, known as an \emph{\ac{IRK}}. When
a device is configured to use a resolvable private address, it generates 22
pseudorandom bits (\emph{prand}), and uses its \emph{local \ac{IRK}} and
\emph{prand} as inputs into a one-way security function, the output of which is
a 24-bit hash~\cite{corespec}. The resolvable private address is created by
setting the most significant bit to 0, second-most significant bit to 1,
concatenated with the 22 bits of \emph{prand}, followed by the 24-bit hash
result. The device then uses this resolvable private address as the source
\ac{MAC} address for a set period of time.  Oftentimes the period is 15
minutes, as~\cite{corespec} recommends 15 minutes as the minimum time interval
between private random address changes. Resolvable private random addresses
have the advantage of allowing potential peers to determine if they already
know a device. If the potential peer has the \ac{IRK} (exchanged during initial
pairing) of the remote device it wishes to connect to, it can determine whether
an advertised random address belongs to that device or not through the
following process: the would-be peer computes the 24-bit hash value given the
\emph{prand} value from the resolvable private address, and the pre-shared
\ac{IRK}. If the value computed locally matches the lower 24 bits of the
resolvable address, the peer's identity has been confirmed to be that
associated with the \ac{IRK} when key exchange was initially done.

\vspace{-4mm}
\subsection{Related Work}
\vspace{-2mm}

Significant previous work exists related to tracking mobile devices via 802.11
Wi-Fi MAC addresses~\cite{Cunche2014, bonne2013wifipi, musa2012tracking,
rajavelsamy2018privacy,o2017mobile,holger_2018,sapiezynski2015tracking},
tracking via cellular
identifiers~\cite{hong2018guti,rajavelsamy2018privacy,o2017mobile,van2015defeating,
mjolsnes2017easy,paget2010practical,strobel2007imsi,ourFirstPaper} and attempting to
correlate randomized 802.11 MAC addresses to the same physical
device~\cite{vanhoef2016mac, martin2017study, rajavelsamy2018privacy}.  By
contrast, our work focuses on the \ac{BLE} technology, and specifically uses
flaws in Apple's Continuity protocol (Section~\ref{sec:continuity}) to
track devices despite randomization of the Bluetooth hardware identifier.
More closely related to this study is previous work on tracking users via
Bluetooth identifiers and the discovery of privacy leakages in Bluetooth
protocols~\cite{haase2004bluetrack,versichele2012use,liebig2012modelling,fawaz,
das2016uncovering,korolova2018cross,becker1tracking}. Of these studies, most
focus on Bluetooth
Classic~\cite{haase2004bluetrack,versichele2012use,liebig2012modelling},
whereas we study Apple's \ac{BLE} Continuity protocol messages exclusively.
Fawaz \etal~\cite{fawaz} develop a tool aimed at preventing privacy leakages in
\ac{BLE} devices by restricting who can discover, scan, and connect to
\ac{BLE}; to date, this tool has not been widely adopted outside of a
laboratory setting. In~\cite{das2016uncovering}, Das \etal examine the privacy
leakages present in wearable fitness tracking devices, as well as the ability
to track these devices and therefore their owners.
Unlike~\cite{das2016uncovering}, Apple devices implement \ac{BLE} MAC address
randomization, which significantly increases the difficulty of tracking them.
Korolova and Sharma~\cite{korolova2018cross} examine the feasibility of
\emph{cross-application tracking} in Android and iOS devices: the ability for
an application to fingerprint applications running on nearby devices by
actively scanning those devices. 

In our work, we do not rely on information gleaned from external applications
installed on our devices, but rather we are able to track users based on
\ac{OS}-default features alone. Like our study, Stute \etal examine a
proprietary protocol used by Apple to enhance interoperability between iOS and
macOS devices. In~\cite{stute2019billion}, Stute \etal reverse engineer Apple's
\ac{AWDL} protocol, an extension to 802.11 that enables AirDrop and
other Apple services. While \ac{AWDL} leverages \ac{BLE} as a discovery
mechanism, the authors are interested in using the \ac{AWDL} implementation for
tracking purposes, rather than \ac{BLE} as in our work.

Contemporaneously and most closely related to our work, Becker
\etal~\cite{becker1tracking} examine tracking devices using randomized \ac{BLE}
identifiers. While they examine \acp{OS} we do not (Windows, Android), their
Apple evaluation considers the \ac{BLE} messages to be largely uninterpretable
data; we exhaustively reverse-engineer the Continuity protocol, revealing both
the structure of the messages as well as what actions are required to produce
them. This affords us the ability to behaviorally profile users, fingerprint
major iOS version and device type, and greatly enhances the potential to track
users despite the use of anonymized MAC addresses, as detailed in
Sections~\ref{sec:analysis} and~\ref{sec:algo_id}.

Finally, an expansive body of literature deals with fingerprinting \ac{OS} type
and version remotely, often by soliciting replies from targets via crafted ICMP
and TCP
messages~\cite{caballero2007fig,chen2014fingerprinting,shamsi2014hershel,
richardson2010limits,beverly2004robust,lyon2009nmap,desmond2008identifying,
cristea2013fingerprinting,kohno2005remote,rye2019sundials}, via DHCP options and
User-Agent strings~\cite{fingerbank}, or in mobile devices by examining
properties of the radio
transmission~\cite{franklin2006passive,neumann2012empirical}, MAC and upper layer
protocols~\cite{cache2006fingerprinting,ellch2006fingerprinting,gentry2016passive},
or both~\cite{xu2015device}. Because the scope
of our study is restricted to Apple devices, our fingerprinting focus is on
differentiating between major release versions of Apple's iOS and macOS
operating systems. Our fingerprinting capability is limited to within \ac{BLE}
transmission distance of the target device, but requires no active
transmissions to elicit a reply from the target and is derived from variations
in the format of Apple's Continuity messages themselves. 

\vspace{-4mm}
\subsection{Apple Continuity}
\label{sec:continuity}
\vspace{-2mm}

\emph{Continuity} is an umbrella term used by Apple to describe interoperability
features between various devices within its ecosystem; for example, the ability
to copy text on an iPhone and paste that same text on a MacBook linked via the
same iCloud account~\cite{macoscontinuity, continuitysupport}. Continuity was
introduced in iOS 8 and OS X Yosemite, although some features require more
recent software versions~\cite{continuityrequirements}. Continuity features
include:
\vspace{4mm}

\begin{itemize}
  \item \textbf{Handoff}, which allows users to start tasks, such as writing an
    email, and continue on another device.
  \item \textbf{Universal Clipboard}, which allows the copying of data from one
    Apple device and pasting it on another.
  \item \textbf{iPhone Cellular Calls}, giving users the ability to make calls
    using their iPhone's cellular connection while on their Mac, iPad, or iPod.
  \item \textbf{Instant Hotspot}, which supports turning an iPhone or iPad into
    a secure hotspot other Apple devices may use without requiring a password.
  \item \textbf{Auto Unlock}, allowing users to unlock a Mac with their Apple
    watch.
  \item \textbf{Continuity Camera}, which transfers photos taken on an iPhone,
    iPad, or iPod touch to a Mac.
\end{itemize}

Continuity features are enabled by the transmission of special \ac{BLE}
advertisement messages sent between devices on the same iCloud account; as
such, all Continuity-enabled devices are \ac{BLE}-capable. In this work, we
reverse-engineer the Apple-proprietary message formats and describe the
operation of those Continuity messages that enable our \ac{OS} fingerprinting
and user tracking techniques in Section~\ref{sec:analysis}.

\section{Methodology}
\label{sec:method}
\vspace{-2mm}

Our analysis was conducted using open-source software and off-the-shelf
commodity hardware.  We perform tests against a wide range of iPhone, iPad,
iPod, airPod, and Apple Watch devices across major and minor iOS versions.
Additionally we experiment with a sample of macOS laptops.  Passive
collection testing was implemented using an Ubuntu environment and an Ubertooth
One USB receiver~\cite{ubertooth} with \emph{2018-12-R1}
firmware~\cite{ubertooth_firmware}.  When utilizing the Ubertooth, we run the
\texttt{ubertooth-btle} software~\cite{ubertooth_btle} to collect \ac{BLE}
advertisement frames.  While running \texttt{ubertooth-btle}, we set the
\texttt{-q} option with \texttt{DLT\_BLUETOOTH\_LE\_LL\_WITH\_PHDR} to ensure
we can capture the \ac{RSSI} value for each frame.  The \texttt{stdout} of
\texttt{ubertooth-btle} is piped directly to Wireshark where we are able to
conduct live analysis. 

As our reverse engineering observations revealed the frame format, message type,
and data attributes of the Apple Continuity protocol, we modified Wireshark's
dissection logic in the \texttt{packet-bthci\_cmd.c} file in order to properly
dissect Continuity messages.  

In section~\ref{sec:active}, we describe a technique to elicit
model-granularity details from Apple devices, and discuss replaying two
previously observed types of Continuity messages. To carry out these active
techniques, we utilize a Sena Technologies UD100 Bluetooth USB adapter, along
with the software \texttt{gatttool} and \texttt{hcitool}. We use
\texttt{gatttool} to query our devices' \ac{GATT}, and \texttt{hcitool} to
spoof arbitrary Continuity frames. 

\vspace{-4mm}
\subsection{Ethical Considerations}
\label{sec:irb}
\vspace{-2mm}

Our collection methodology is entirely passive.  At no time did we attempt to
decrypt any user data, alter normal network behavior, or attempt to track any
individuals not associated with our research team without their prior
knowledge. Additionally, in order to evaluate the privacy flaws we present in
this work, we conduct a variety of experiments on lab devices owned by the
authors and the authors' institutions.  These devices were allowed to
communicate with legitimate network services.  Given the nature of our data
collection, we consulted with our \ac{IRB}.

The primary concerns of the \ac{IRB} centered on: i) the information collected;
and ii) whether the experiment collects data ``about whom'' or ``about what.''
Because we limit our analysis to \ac{BLE} advertisement frames we do not
observe \ac{PII}.  Further, humans are incidental to our experimentation as our
interest is in \ac{OS} and hardware profiling, device usage, and the analysis
of the randomization of \ac{BLE} device layer-2 \ac{MAC} addresses, or
``what.'' 

Finally, in consideration of beneficence and respect for persons, our work
presents no expectation of harm, while the concomitant opportunity for network
measurement and security provides a societal benefit.  Our experiment was
therefore determined by the \ac{IRB} to not be human subject research.

\vspace{-6mm}
\section{Analysis}
\label{sec:analysis}
\vspace{-2mm}

This section is divided into three parts:

\bi
   \item Analysis of passively collected data to reverse engineer the frame
format and data attributes of Apple's \ac{BLE} Continuity framework.  We show
through this effort, that Continuity features leak a significant amount of data
related to user behavior.
   \item Active attacks transmitting tailored \ac{BLE} frames to elicit
responses from Apple devices revealing further details regarding user information
and behavior.
  \item A comprehensive evaluation of the effectiveness of these attacks with
respect to an adversary's ability to identify and track devices and users.
\ei

\vspace{-2mm}

Since we analyze many different types of Continuity messages, each with
different flaws, we organize our findings by calling attention specifically to
the following categories, based on what information is leaked:

\vspace{2mm}
\bi
  \item \textbf{OS fingerprinting
  \item Device fingerprinting
  \item Tracking
  \item User / Device Activity
  \item Device attributes}
\ei

\vspace{-2mm}
When a flaw is observed, we classify it into one or more of these categories
and describe how it can be exploited by a potential adversary.
\vspace{-2mm}

\vspace{-4mm}
\subsection{Passive Analysis Reverse Engineering}
\label{sec:passive}
\vspace{-2mm}

We evaluate Apple's proprietary Continuity protocol by inspecting
\ac{BLE} frames emitted from iOS and macOS devices across Apple's
ecosystem and \ac{OS} versions.

%JM:no longer on right page :( not sure i can fix this 
\begin{figure*}[!t]
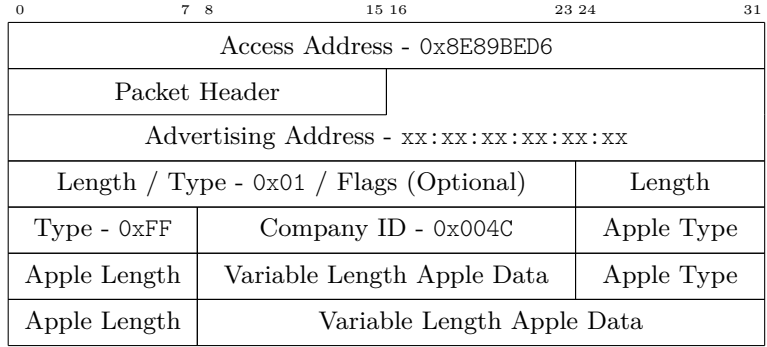

\centering
  \begin{bytefield}{32}
    \bitheader{0,7,8,15,16,23,24,31} \\
    \wordbox{1}{Access Address - \texttt{0x8E89BED6}} \\
    \bitbox{16}{Packet Header} & \bitbox[lrt]{16}{} \\
    \wordbox[lrb]{1}{Advertising Address - \texttt{xx:xx:xx:xx:xx:xx}} \\
    \bitbox{24}{Length / Type - \texttt{0x01} / Flags (Optional)} & \bitbox{8}{Length} \\
    \bitbox{8}{Type - \texttt{0xFF}} & \bitbox{16}{Company ID - \texttt{0x004C}} & \bitbox{8}{Apple Type} \\
    \bitbox{8}{Apple Length} & \bitbox{16}{Variable Length Apple Data} & \bitbox{8}{Apple Type} \\
    \bitbox{8}{Apple Length} & \bitbox{24}{Variable Length Apple Data} \\
  \end{bytefield}
\vspace{-3mm}
  \caption{Apple BLE Frame Format with Hardcoded Field Values}
  \label{fig:appleFrame}
\end{figure*}

As described in Section~\ref{sec:method}, we collect \ac{BLE} frames passively
using an Ubertooth with \texttt{stdout} piped to Wireshark (using our custom
dissector), allowing for real-time dynamic analysis via a live capture display.
Apple Continuity frames are transmitted on all three \ac{BLE} advertisement
channels; as such, our Ubertooth collection setup requires monitoring
only a single advertisement channel.

\subsubsection{BLE Advertisement Frame Prevalence}
\label{sec:prevalence}
\vspace{-2mm}

As a preliminary study, we first sought to understand the prevalence of
\ac{BLE} in representative public locations.  Our goal was to understand what
types of devices, \acp{OS}, and ecosystems were employing \ac{BLE} and whether
privacy countermeasures such as \ac{MAC} address randomization were frequently
deployed.  Table~\ref{table:advertStats} depicts two distinct measurements: the
first is a multi-hour single collection, while the second comprises several
distinct collections over two days.  In both cases our results indicate that
only two major ecosystems are commonly observed utilizing \ac{BLE} \ac{MAC}
address randomization -- Apple and Microsoft Windows devices. It should be
noted that the counts for \emph{Random} \ac{MAC} addresses are skewed, and
therefore should not be interpreted as the number of distinct devices observed.
This is a reflection of the interval policy used by Apple and Microsoft, in
which the random \ac{BLE} \ac{MAC} address rotates every 15 minutes.

\begin{table}[!htb]
  \center
  \caption{Advertisement Frames}
  \label{table:advertStats}
  \begin{tabular}{r|c|c|c}
    \toprule
     \multicolumn{2}{c}{} & Test 1 & Test 2 \\
    \midrule  
      \multicolumn{2}{c}{} & \multicolumn{2}{c}{Count} \\
      \multirow{2}{*}{Address Type} & Public & 26 & 57 \\
      & Random & 726 & 1,518 \\
    \midrule
     \multirow{5}{*}{Company ID$\dagger$} & Apple & 692 & 1296 \\
      & Microsoft   & 30 & 201 \\
      & Garmin      & 2 & 9 \\
      & Samsung     & 0 & 3 \\
      & All Others  & 2 & 9 \\
    \midrule
      \multicolumn{4}{l}{$\dagger$ Randomized Devices Only} \\
    \bottomrule
  \end{tabular}
\end{table}

Microsoft's Connected Devices Platform (MS-CDP) discovery protocol~\cite{mscdp}
provides a framework to allow users to verify and authenticate devices and
exchange messages between devices.  As the protocol is delineated in a
published specification we center our analysis on reverse engineering Apple's
Continuity protocol. To refine our analysis to focus solely on Apple \ac{BLE}
traffic, we filter for \ac{BLE} frames containing Apple's Company ID
(0x004C)~\cite{sig}.

\centerline{\textbf{Device Fingerprinting}}
While the Company ID is required as per specification~\cite{sig}, it allows for
simple identification of \ac{BLE} traffic generated by Apple devices. 

\vspace{-4mm}
\subsubsection{Configuration Settings - Disabling Continuity}
\vspace{-2mm}

An underlying condition for Continuity messages to be sent is the user having
their device associated with an iCloud account to which at least two devices are
registered; because users regularly neglect to remove old Apple products from
their iCloud account years after discarding or retiring the device, this
condition is routinely met even when the user may only actively use one Apple
device. 

While they are enabled by default, Handoff messages described in Section 
\ref{sec:handoff} are unique in that they can
be explicitly turned off in the Settings Menu. Some message types require a
device with a cellular connection to be associated with the iCloud account in
order to be generated by the user, namely the WiFi Settings and Instant Hotspot
messages, outlined respectively in Sections~\ref{sec:wifisettings} and~\ref{sec:hotspot},
which need a cellular-capable device to act as a hotspot.

Activating Airplane Mode does not disable the transmission of Continuity
messages in either iOS 11 or 12, regardless of whether Airplane Mode is
activated from the Control Center or Settings Menu. Similarly, disabling
Bluetooth from the Control Center in iOS 11 or 12 does not discontinue the
transmission of Continuity messages~\cite{controlCenter}. We determined
that the only way to stop transmission of all Continuity messages 
is to disable Bluetooth from the Settings Menu.

\begin{table*}[!ht]
\vspace{-3mm}
\centering
  \caption{Most Commonly Observed Continuity Messages}
  \label{table:messages}
  \vspace{-2mm}
  \begin{tabular}{r|c||c|c|c|c|c|c|c|c|c|c}
    \toprule
      & & \multicolumn{5}{c}{iOS Version}  & \multicolumn{5}{c}{Vulnerability} \\
      Type & Value & 8 & 9 & 10 & 11 & 12 &
       OS FP & Device FP & Tracking & Activity &
       Attributes
      \\
    \midrule
      Watch Connection & 11 & N & N & N & Y & Y & \xmark & \checkmark & \xmark & \xmark& \xmark\\
      Handoff & 12 & Y & Y & Y & Y & Y & \xmark & \xmark & \checkmark &\checkmark & \checkmark\\
      Wi-Fi Settings & 13 & 8.1+ & Y & Y & Y & Y & \xmark & \xmark & \checkmark & \checkmark & \xmark \\
      Instant Hotspot & 14 & 8.1+ & Y & Y & Y & Y & \xmark & \xmark &\checkmark & \checkmark & \checkmark\\
      Wi-Fi Join Network & 15 & N & N & N & Y & Y & \xmark & \xmark & \checkmark
      &\checkmark & \xmark \\
      Nearby & 16 & N & N & Y & Y & Y & \checkmark & \xmark & \checkmark & \checkmark & \checkmark \\
    \bottomrule
  \end{tabular}
  \vspace{-3mm}
\end{table*}

\vspace{-4mm}
\subsubsection{Overall Message Structure}
\label{sec:messaging}
\vspace{-2mm}

After segregating all Apple traffic from our collection by filtering for
Apple's Company ID, we observe that Apple's Continuity frames adhere to a
simple Type-Length-Value (TLV) structure delineated in
Figure~\ref{fig:appleFrame}.  Of note, multiple Continuity message types are
often concatenated together in this TLV format, allowing them to be passed in a
single advertisement frame. 

\centerline{\textbf{Device Fingerprinting}}
We observe that optional \ac{BLE} advertisement flags indicate a device
category, allowing us to delineate MacBooks from mobile devices
(iPhone, iPad, iPod, and watches). Specifically, we investigate the following flags:

\bi
  \item Simultaneous LE and BR/EDR to Same Device Capable Host (\textbf{H})
  \item Simultaneous LE and BR/EDR to Same Device Capable Controller (\textbf{C})
  \item Peripheral device is LE only (\textbf{LE})
\ei

\vspace{-3mm}
Mobile devices were observed with flags \textbf{H, C,} and \textbf{LE} set to
\texttt{1,1,0}, whereas MacBooks were set to \texttt{0,0,1}.  AirPods lacked any
flags and were thereby easily identifiable as the only device type with no flag
attributes. 

After adding the identified TLV structure, our custom-defined Apple \ac{BLE}
fields, and associated attributes to the \texttt{packet-bthci\_cmd.c} Wireshark
dissector, we proceed to reverse engineer the most commonly observed Continuity
messages.  For each new message type and attribute we update the dissector,
recompile, and reevaluate across multiple devices and \acp{OS}.
Table~\ref{table:messages} highlights the message types and corresponding
values we observed.  Also annotated are the message type mappings to applicable
iOS versions.  The message types' descriptive names were generated by our group in
an attempt to properly categorize the message.

\vspace{-5mm}
\subsubsection{Other} 
\vspace{-3mm}
While iBeacon messages uniquely identify iBeacon nodes, these devices, as their
name implies are not meant to be anonymous and we conduct no further analysis
of the iBeacon message type.

Additionally, AirDrop and AirPlay messages were observed, albeit infrequently
and were not examined in our study; \cite{stute2019billion} \etal provide a
thorough investigation of privacy leakages and tracking mechanisms enabled by
the AirDrop protocol. Though we did not make direct use of any of these message
types, we reverse engineered enough of their format to allow us to detect and
ignore them in our study.

\centerline{\textbf{Device Fingerprinting}}
Lastly, we observed that AirPods transmit a unique message type and were
trivially identified via observation of these messages.

\vspace{-5mm}
\subsubsection{Watch Connection}
\vspace{-3mm}

An Apple Watch transmits message type \texttt{11} when the watch has lost a
Bluetooth connection with the paired iPhone. 

\centerline{\textbf{Device Fingerprinting}}
This message distinctly identifies Apple Watches as they are solely transmitted
from an Apple Watch.

\vspace{-5mm}
\subsubsection{Handoff}
\label{sec:handoff}
\vspace{-3mm}

Handoff messages occur when a user interacts with a Handoff-enabled application
such as Mail, Maps, Safari, Calendar, Contacts, Pages, Numbers, Keynote, and
third-party applications such as Airbnb and Google Chrome~\cite{handoffApps}.
In addition to being triggered by user interaction with Continuity-enabled
applications, Handoff messages are also observed when these applications are
opened or closed.  Unlike other message types, Handoff messages can be disabled
explicitly through the settings page, though they are enabled by default.
Handoff messages are also only observed when tied to an iCloud account that
contains two or more Apple devices, as Handoff cannot work with a lone device.

\begin{figure}[!h]
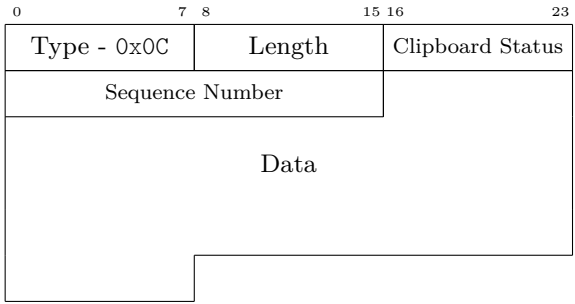

  \centering
  \begin{bytefield}{24}
    \bitheader{0,7,8,15,16,23} \\
    \bitbox{8}{Type - \texttt{0x0C}} & \bitbox{8}{Length} & \bitbox{8}{\small Clipboard Status} \\
    \bitbox{16}{\small Sequence Number} & \bitbox[lrt]{8}{} \\
    \wordbox[lr]{2}{Data} \\
    \bitbox[l]{8}{} & \bitbox[br]{16}{} \\
    \bitbox[lrb]{8}{}\\
  \end{bytefield}
  \vspace{-4mm}
  \caption{Handoff Message - Frame Format}
  \label{fig:handoffFrame}
  \vspace{-2mm}
\end{figure}

Having identified the user behaviors that generate Handoff messages, we focus on
recovering privacy-sensitive information with the message itself.
Figure~\ref{fig:handoffFrame} depicts the Handoff frame, indicated by a Type
field of \texttt{0x0C}.

\vspace{2mm}
\centerline{\textbf{Device Attributes}}
The Handoff message contains a one-byte field that contains an indicator we
call ``clipboard status'' indicating when a user has recently copied data to
the Universal Clipboard and is now available to transfer to nearby Apple
devices.  This behavior was observed across iOS 10, 11, and 12.  A value of
0x08 indicates that data are stored in the Universal Clipboard, and 0x00 if
not. 

\centerline{\textbf{Tracking}} 
A two-byte sequence number follows the clipboard status field.  The sequence
number increments only when a new action occurs in a Handoff-enabled
application, the application is opened/closed, the phone is unlocked, or the
phone is rebooted.  As such, the sequence number increments monotonically,
slowly, and at a rate proportional to Handoff-enabled app use, allowing for
long-duration tracking. The remaining bytes in the frame appear to be encrypted
data and provide no specific details about the user's activity that we could
infer.

Importantly, sequence numbers are not affected by \ac{MAC} address
randomization.  We observe \ac{MAC} address changes that preserve the sequence
number and encrypted Handoff data before and after the expiration of the
\emph{private\_addr\_int} timer. This allows for the trivial association of two
random \ac{MAC} addresses, and, because addresses change on a fixed 15 minute
schedule, allows a passive adversary to prepare for the next rotation 15 minutes
hence.  This behavior was consistently observed across all iOS major and minor
versions we tested, from iOS 9 through 12.3. We describe this tracking
vulnerability in detail in Section~\ref{sec:algo_id}.

\centerline{\textbf{User Activity}}
The sequence number also carries information about private user behavior. Since
it increases only in response to user-generated events, it serves as a crude,
real-time measurement of user activity. Measuring the sequence number of a
device at two different times allows an adversary to infer how much the phone
was used during that time period, leaking information about the activity
between measurements. 

\subsubsection{Wi-Fi Settings}
\label{sec:wifisettings}
\vspace{-1mm}

\begin{figure}[!htb]
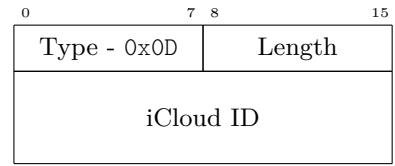

  \centering
  \begin{bytefield}{16}
    \bitheader{0,7,8,15} \\
    \bitbox{8}{Type - \texttt{0x0D}} & \bitbox{8}{Length} \\
    \wordbox{2}{iCloud ID} \\
  \end{bytefield}
  \vspace{-3mm}
  \caption{Wi-Fi Settings Message - Frame Format}
  \label{fig:wifiSettingsFrame}
\end{figure}

Another Continuity message type, which we refer to as ``Wi-Fi Settings'', is
transmitted when the user navigates to the Wi-Fi Settings page in iOS or clicks
on the Wi-Fi and network status icon on the top of their screen in macOS. While
the settings page is open, an Apple device will continuously transmit Wi-Fi
Settings frames, as depicted in Figure~\ref{fig:wifiSettingsFrame}.  This
feature requires that a device other than the one in use is registered to the
same iCloud account, and that the second device has a cellular radio and is
Instant Hotspot capable.

\vspace{2mm}
%WiFi Settings

\centerline{\textbf{User Activity}}
This message indicates to an observer that the user is currently on the Wi-Fi
Settings page and is likely configuring or about to connect to a Wi-Fi
network.

\centerline{\textbf{Tracking}} 
A four-byte data field representing an iCloud-derived ID trivially links all
other devices tied to the same iCloud account.  The derived ID is rotated on a
24 hour basis for all devices on the account where each device remains synchronized
with all other devices on the same account; as the ID is ephemeral, however, it
is not trackable for more than 24 hours.

\vspace{-3mm}
\subsubsection{Instant Hotspot}
\label{sec:hotspot}
\vspace{-1mm}

%JM: yucky sentence
Instant Hotspot is an Apple Continuity feature that allows users to share
cellular network connectivity among iCloud-linked devices by creating a bridged
Wi-Fi connection.  The Instant Hotspot feature allows Apple devices to
seamlessly identify and connect to these personal hotspot-enabled devices
through the use of the Instant Hotspot message.  Devices configured to support
Instant Hotspot will automatically begin transmitting Instant Hotspot messages
when another device on the same iCloud account is nearby and is transmitting
Wi-Fi Settings messages containing the correct iCloud-derived ID.  The
hotspot-enabled device continues to broadcast Instant Hotspot messages as long
as the other device remains on the Wi-Fi Settings menu.

\vspace{3mm}
\centerline{\textbf{Device Activity}} 
This message is inherently descriptive of the device's ability to support
acting as an \ac{AP}, and assuming a connection is established, the device
will be recognizable in the 802.11 domain as an \ac{AP}. 

\vspace{-2mm}
\begin{figure}[!htb]
  \hspace{-2.4cm}
%\begin{minipage}{.5\textwidth}
  \begin{tikzpicture}[font=\sffamily,>=stealth',
                    thick, 
                    commentl/.style={text width=3.8cm, align=right},
                    commentr/.style={commentl, align=right},]

    \node[] (client) {Client};
    \node[right=3.5cm of client] (ap) {Instant Hotspot};

    \node[below left = 1mm and 2mm of client.south, commentl]
         {\textbf{Wi-Fi Settings}\\[-1.5mm]{\textbf{Page}}};

    \draw[->] ([yshift=-.7cm]client.south) coordinate (0x0d) -- 
              ([yshift=-.7cm]0x0d-|ap) coordinate (0x0dEnd) node[pos=.5, above, sloped]
              {\texttt{Wi-Fi Settings (0x0D)}};

    \draw[->] ([yshift=-.1cm]0x0d.south) coordinate (a) -- 
              ([yshift=-.7cm]a-|ap) coordinate (b) ;

    \draw[->] ([yshift=-.1cm]a.south) coordinate (aa) -- 
              ([yshift=-.7cm]aa-|ap) coordinate (bb) ;
  
    \draw[dotted] (client.255)--([yshift=-4mm]client.255|-0x0d);

    \draw[->] ([yshift=-.4cm]0x0dEnd) coordinate (0x0e) -- 
              ([yshift=-.7cm]0x0e-|client) coordinate (0x0eEnd) node[pos=.5, above, sloped] 
              {\texttt{Instant Hotspot (0x0E)}};

    \draw[->] ([yshift=-.1cm]0x0e.south) coordinate (c) -- 
              ([yshift=-.7cm]c-|client) coordinate (d) ;

    \draw[->] ([yshift=-.2cm]0x0e.south) coordinate (cc) -- 
              ([yshift=-.7cm]cc-|client) coordinate (dd) ;

    \node[below left = 2mm and 2mm of 0x0eEnd.south, commentl]
         {\textbf{User Chooses}\\[-1.5mm]{\textbf{Hotspot}}};

    \draw[->] ([yshift=-.7cm]0x0eEnd-|client) coordinate (preq) -- 
              ([yshift=-.7cm]preq-|ap) coordinate (presp) node[pos=.5, above, sloped] 
              {Probe Request};

    \draw[dotted] ([yshift=1mm]client.255|-0x0eEnd) --
                  ([yshift=-1mm]client.255|-preq);

    \draw[->] ([yshift=-.4cm]presp) coordinate (found) -- 
              ([yshift=-.7cm]found-|client) coordinate (nowconnect) node[pos=.5, above, sloped] 
              {Probe Response};

    \draw[->] ([yshift=-.3cm]nowconnect-|client) coordinate (auth) -- 
              ([yshift=-.7cm]auth-|ap) coordinate (ack) node[pos=.5, above, sloped] 
              {Authentication Request};

    \draw[thick, shorten >=-1mm] (client) -- (client|-ack);
    \draw[thick, shorten >=-1mm] (ap) -- (ap|-ack);

  \end{tikzpicture}
%  \captionsetup{justification=RaggedLeft}
  \caption{Instant Hotspot Discovery and Connection Setup}
  \label{fig:a}
%  \end{minipage} 
%\hspace{1.9cm}
\end{figure}
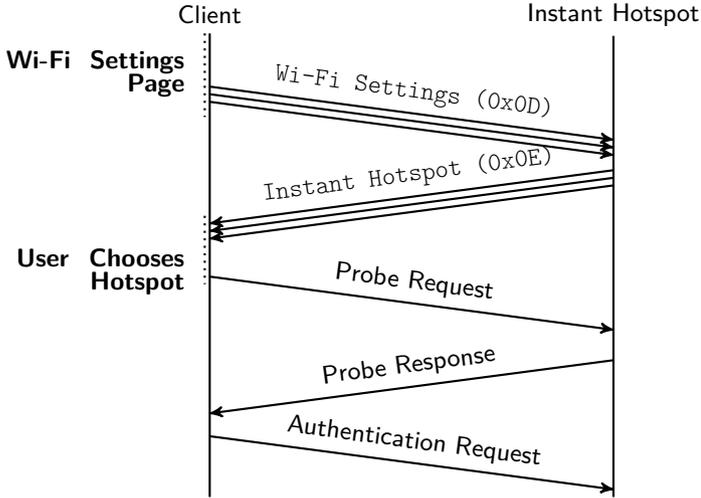

\vspace{-3mm}
\centerline{\textbf{Tracking}} 
The Instant Hotspot connection process, described in an abbreviated form in
Figure~\ref{fig:a}, highlights several tracking issues.  First, upon attempting
to connect to an Instant Hotspot network, the client searches for the Hotspot
by sending directed and broadcast probe request frames.  This is similar to the
well-documented probe request privacy flaw in which a user can be profiled
based on the \ac{SSID} of the networks for which it actively
searches~\cite{barbera2013signals, cunche2012know}.  

Upon receiving a probe request from a would-be client, the Instant Hotspot
device transmits probe responses and beacon frames using a
deterministically-generated, locally-administered \ac{MAC} address
rather than its true, \emph{global public} \ac{MAC} address. The probe
responses and beacons transmitted by the Instant Hotspot device provide a
second tracking mechanism through the inclusion of a vendor-specific \ac{IE}
that includes a reversible permutation of the Bluetooth and Wi-Fi
globally-unique \ac{MAC} addresses as shown in~\cite{martin2017study}. 

Finally, the client device transmits an 802.11 authentication frame, in which
the source address is the \emph{global public} \ac{MAC} address of the device.
Strangely, we observe that probe request frames also use the \emph{global public}
\ac{MAC} address of the device; this is unusual because iOS devices generally
randomize \ac{MAC} addresses in probe requests when in an un-associated state. 

\begin{figure}[!htb]
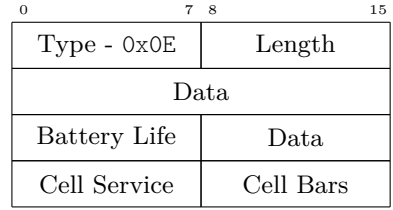

  \centering
  \begin{bytefield}{16}
    \bitheader{0,7,8,15} \\
    \bitbox{8}{Type - \texttt{0x0E}} & \bitbox{8}{Length} \\
    \bitbox{16}{Data} \\
    \bitbox{8}{Battery Life} & \bitbox{8}{Data} \\
    \bitbox{8}{Cell Service} & \bitbox{8}{Cell Bars} \\
  \end{bytefield}
  \vspace{-5mm}
  \caption{Instant Hotspot Message - Frame Format}
  \label{fig:instantHotspotFrame}
  \vspace{-5mm}
\end{figure}

\centerline{\textbf{Device Attributes}} 
The Instant Hotspot message reveals surprising device characteristics in
plaintext as shown in Figure~\ref{fig:instantHotspotFrame}.  
Specifically, the device's battery life, cellular service type (e.g. LTE, 3G,
EV-DO), and cellular service quality (measured as a function of number of bars)
are all transmitted unencrypted.

% Instant Hotspot

\begin{comment}
\begin{table}[!htb]
  \center
  \caption{Cellular Service Codes}
  \label{table:cellCodes}
  \begin{tabular}{r|l}
    \toprule
      Value & Cellular Type \\ 
    \midrule
     0 & No Icon $\dagger$ \\
     1 & 1xRTT \\
     2 & GPRS \\
     3 & EDGE \\
     4 & 3G (EV-DO) \\
     5 & 3G \\
     6 & 4G \\
     7 & LTE \\ 
    \midrule
    \multicolumn{2}{r}{$\dagger$: Degraded Signal} \\
   \bottomrule
  \end{tabular}
\end{table}
\end{comment}

\centerline{\textbf{Tracking}} 
Lastly, due to the nature of how an Instant Hotspot message is elicited we can
trivially determine that the initiator device (Wi-Fi Settings message transmitter) 
and the Instant Hotspot device are associated to the same iCloud
account. 

\vspace{-6mm}
\subsubsection{Wi-Fi Join Network}
\vspace{-3mm}

An additional Wi-Fi themed Continuity message type, in which we annotate as
``Wi-Fi Join Network'', is transmitted when a user attempts to connect to an
encrypted network from the Wi-Fi Settings page.  We call attention to the fact
that this message is only sent when a password is required; therefore, open and
captive portal-enabled networks will not generate Wi-Fi Join Network messages. 

\centerline{\textbf{User / Device Activity}} 
We note that the observation of a Wi-Fi Join message indicates the intent by a
user to connect to an encrypted network regardless of whether the proper
credentials are entered.  

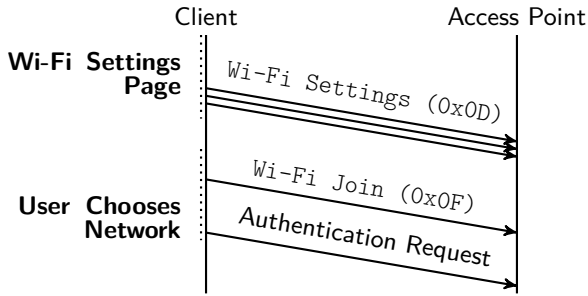
\begin{figure}[!htb]
%  \begin{minipage}{.45\textwidth}
  \hspace{-.7cm}
  \begin{tikzpicture}[font=\sffamily,>=stealth',
                    thick, 
                    commentl/.style={text width=3cm, align=right},
                    commentr/.style={commentl, align=left},]

    \node[] (client) {Client};
    \node[right=2.5cm of client] (ap) {Access Point};

    \node[below left = 1mm and 2mm of client.south, commentl]
         {\textbf{Wi-Fi Settings}\\[-1.5mm]{\textbf{Page}}};

    \draw[->] ([yshift=-.7cm]client.south) coordinate (0x0d) -- 
              ([yshift=-.7cm]0x0d-|ap) coordinate (0x0dEnd) node[pos=.5, above, sloped]
              {\texttt{Wi-Fi Settings (0x0D)}};

    \draw[->] ([yshift=-.1cm]0x0d.south) coordinate (a) -- 
              ([yshift=-.7cm]a-|ap) coordinate (b) ;

    \draw[->] ([yshift=-.1cm]a.south) coordinate (aa) -- 
              ([yshift=-.7cm]aa-|ap) coordinate (bb) ;
  
    \draw[dotted] (client.255)--([yshift=-4mm]client.255|-0x0d);

    \draw[->] ([yshift=-.3cm]bb-|client) coordinate (wifijoin) -- 
              ([yshift=-.7cm]wifijoin-|ap) node[pos=.5, above, sloped] 
              {\texttt{Wi-Fi Join (0x0F)}};

    \draw[->] ([yshift=-.7cm]wifijoin-|client) coordinate (auth) -- 
              ([yshift=-.7cm]auth-|ap) coordinate (connect) node[pos=.5, above, sloped] 
              {Authentication Request};

    \node[below left = 1mm and 2mm of wifijoin.south, commentl]
         {\textbf{User Chooses}\\[-1.5mm]{\textbf{Network}}};

    \draw[dotted] ([yshift=1mm]client.255|-bb) --
                  ([yshift=-1mm]client.255|-auth);

    \draw[thick, shorten >=-1mm] (client) -- (client|-connect);
    \draw[thick, shorten >=-1mm] (ap) -- (ap|-connect);

  \end{tikzpicture}
  \caption{Wi-Fi Join -- Authentication Frame Global \ac{MAC} Exposed}
  \label{fig:b}
 \vspace{-4mm}
%\end{minipage}
%  \label{fig:blewifi}
\end{figure}

\centerline{\textbf{Tracking}} 
A similar tracking flaw to the one described in Section~\ref{sec:hotspot}
allows an adversary to link the \ac{BLE} communication (Wi-Fi Join message) to
the frames observed in the 802.11 (Wi-Fi) domain.  Specifically, as delineated
in Figure~\ref{fig:b} the observer can compare the timestamps of the Wi-Fi Join
message to the authentication frames collected at the same time.  As the
authentication frame contains the \emph{global public} \ac{MAC}
address~\cite{martin2017study}, anonymization is broken in both the Wi-Fi and
\ac{BLE} protocols.

Due to the state in which this message is sent in the process (prior to
authentication), we can retrieve the client's \emph{global public} \ac{MAC}
address even when a user attempts to authenticate with an invalid passcode.

\begin{figure}[!htb]
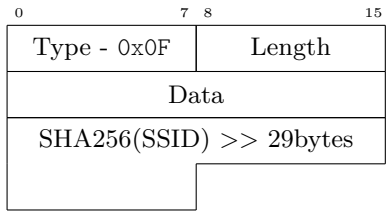

  \centering
  \begin{bytefield}{16}
    \bitheader{0,7,8,15} \\
    \bitbox{8}{Type - \texttt{0x0F}} & \bitbox{8}{Length} \\
    \wordbox[lrb]{1}{Data} \\
    \bitbox[lrt]{16}{SHA256(SSID) $>>$ 29bytes } \\
    \bitbox[lrb]{8}{} & \bitbox[t]{8}{}
  \end{bytefield}
  \caption{Wi-Fi Join Network Message - Frame Format}
  \label{fig:wifiJoinFrame}
  \vspace{-8mm}
\end{figure}

Another glaring implementation flaw centers on a three-byte \ac{SSID} field in
the Wi-Fi Join frame, depicted in Figure~\ref{fig:wifiJoinFrame}.  This field
represents the first three bytes of a SHA256 hash of an \ac{SSID} the client
device is attempting to join.  An adversary can pre-compute hashes of
nearby \acp{SSID}, allowing for trivial correlation. 

\subsubsection{Nearby}
\label{sec:nearby}
\vspace{-2mm}

Nearby messages, presumably intended to keep \emph{nearby} devices aware of the
state of other devices in the same iCloud ecosystem, are transmitted frequently
by Continuity-enabled Apple devices.  As of iOS 12 and macOS Mojave, Nearby
messages \textbf{never} stop transmitting, and are sent at a rate of over 200
frames per minute. Because of the frequency and consistency with which Nearby
messages are transmitted and the user-behavioral data contained in their
payload, Nearby messages represent a serious privacy and tracking issue.  
%JM:clean up claim of never, to equal below clarification, or is this okay?

Support for Nearby messages began with release of iOS 10. Observable changes in
the frame format and usage correspond with each iOS major version release
thereafter.  In earlier implementations (iOS 10 and 11, and macOS High Sierra)
Nearby messages timeout after a period of inactivity, meaning that devices cease to
transmit Nearby messages when a device is left inactive after approximately 30
seconds. As of iOS 12 and macOS Mojave, devices continuously transmit Nearby
messages while \ac{BLE} services are not disabled and the device is on, and so 
long as the lid is not closed on a macOS device. 

% Nearby 
\begin{figure}[!htb]
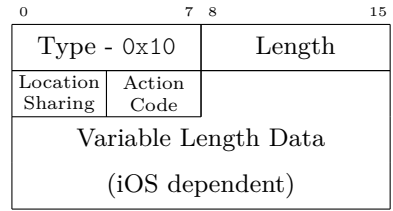

  \centering
  \begin{bytefield}{16}
    \bitheader{0,7,8,15} \\
    \bitbox{8}{Type - \texttt{0x10}} & \bitbox{8}{Length} \\
    \bitbox{4}{\scriptsize{Location Sharing}} & \bitbox{4}{\scriptsize{Action Code}} 
    & \bitbox[lrt]{8}{} \\
    \wordbox[lr]{1}{Variable Length Data} \\
    \wordbox[lrb]{1}{(iOS dependent)} \\
  \end{bytefield}
  \vspace{-4mm}
  \caption{Nearby Message - Frame Format}
  \label{fig:nearbyFrame}
  \vspace{-4mm}
\end{figure}

\centerline{\textbf{User / Device Activity}} 
The Nearby message contains a one-byte field, referenced in
Figure~\ref{fig:nearbyFrame}, of which the least significant nibble we
designate the ``Action Code'' field, as it indicates the \emph{action} or
\emph{state} of the Apple device.  The most significant nibble, not observed in
use prior to iOS 12.3, indicates if the device has been configured to ``Share
My Location'' with family and friends.  As of iOS 12.3 this field
is set to \texttt{1} when the user specifically selects this device as the
sharer of the locational data. Every iCloud account has a single device
selected as its location sharer, and this field is set to \texttt{1} for this
selected device, and \texttt{0} for all other devices on the account.  This
behavior is observed regardless of whether location services is on.

Our research highlighted in Table~\ref{table:actionCodes} describes seven
commonly observed Action Codes.

\vspace{-4mm}
\begin{table}[!ht]
  \center
  \caption{Action Codes}
  \label{table:actionCodes}
  \vspace{-3mm}
  \begin{tabular}{r|l}
    \toprule
      Type & Description \\ 
    \midrule
      1    & iOS recently updated \\
      3    & Locked Screen \\
      7    & Transition Phase \\
      10   & Locked Screen, Inform Apple Watch \\
      11   & Active User \\
      13   & Unknown \\
      14   & Phone Call or Facetime\\
    \bottomrule
  \end{tabular}
  \vspace{-5mm}
\end{table}

\begin{description}

 \item [Action Code 11] A user is actively interacting with the associated
device (iOS or macOS).  This Action Code will continue to be transmitted until
a period of $\sim$30 seconds of inactivity, upon which it will then send Nearby
messages with Action Code 7, and later Action Code 3 as the screen becomes
locked (iOS). 

 \item [Action Code 3] An iOS device in a screen locked state, 
indicating a lack of interaction between the user and the device. 

 \item [Action Code 10] Informs a paired Apple Watch that the connected mobile device is in
a locked screen state.  In this scenario the mobile device will send only
Action Code 10 messages vice the previously mentioned Action Code 3, likely to 
indicate that the phone will forward notifications to the watch when the
device is locked and that the watch should display the notifications. 

 \item [Action Code 7] Observed after $\sim$30 seconds of user-device (iOS or
macOS) inactivity or when the device transitions from a locked state to that of
an active state (caused by waking the device). 

 \item [Action Code 14] Observed after the release of iOS 12.3 when transmitted from iOS
devices in an active phone call or Facetime session.

 \item [Action Code 13] Observed in the wild after the release of iOS 12.3, however we were
unable to reproduce this message in our laboratory experiments.

 \item [Action Code 1] We observed this Action Code rarely. In our experiments,
we observed this Action Code from iOS devices recently updated to a new iOS
version which have not been rebooted after the full update.  Additionally, we
often observed these from locked macOS devices, however we were unable to
definitively attribute an action to these messages.  

\end{description}

\vspace{-4mm}

\centerline{\textbf{Device Attributes}} 
iOS 12 and macOS Mojave Nearby messages contain additional data, highlighted in
Table~\ref{table:n_messages}, in which the state of the Wi-Fi radio can be
inferred, namely whether Wi-Fi is enabled or disabled.  If the first byte of
the data field is \texttt{0x18} the Wi-Fi radio is off, whereas when the value
is \texttt{0x1C} the Wi-Fi radio is on. 

\vspace{-4mm}
\begin{table}[!htb]
  \center
  \caption{Nearby Message (iOS) -- Data Field}
  \label{table:n_messages}
  \vspace{-2mm}
  \begin{tabular}{r|c|c|c}
    \toprule
      \multirow{2}{*}{Feature} & \multicolumn{3}{c}{iOS Version} \\
      & 10 & 11 & 12 \\ % & macOS\\
    \midrule
      Length (bytes) & 1 & 4 & 4 \\ % & 1\\
      Byte 1   & 0x00 & 0x10 & 0x18 || 0x1C \\ %& 0x08 \\
      Byte 2-4 &  -   & Data & Data \\ %& - \\
    \midrule
    \multicolumn{3}{r}{$\dagger$: 0x18 (Wi-Fi Off), 0x1C (Wi-Fi On)} & \\
    \bottomrule
  \end{tabular}
  \vspace{-6mm}
\end{table}

\centerline{\textbf{OS Fingerprinting}} 
The variable nature of the Nearby message data field allows for \ac{OS} version
fingerprinting.  Specifically, the length of the frame and the value of the
first byte distinctly reveal the iOS major version.  As such we can infer a
device's iOS major version as either iOS 10, 11, or 12 with 100\% accuracy.
Similarly, we can identify macOS as Mojave or pre-Mojave \ac{OS} versions,
however we must first evaluate the \ac{BLE} flags as described in
section~\ref{sec:messaging} in order to separate iOS from macOS devices.

\centerline{\textbf{Tracking}} Similar to the privacy flaw discussed in
Section~\ref{sec:handoff}, the Nearby data field does not immediately
change when the \ac{MAC} address is periodically rotated, allowing for
trivial tracking of a device across \ac{MAC} address changes. This is further
compounded with the behavior observed in iOS 12 and macOS Mojave where Nearby
messages are continuously transmitted.

\subsection{macOS \ac{BLE} \ac{MAC} Randomization Breaks Itself}
\label{sec:macOS}
\vspace{-3mm}

For macOS devices running High Sierra and Mojave, we observed that Nearby
messages change when Handoff messages are being sent concurrently. Normally, all
Apple Continuity messages utilize the current \emph{resolvable private random address}.
This remains true with macOS under circumstances where only Nearby messages are
being transmitted by the macOS device.  However, when a Handoff message or
Wi-Fi Settings message is sent, the Nearby messages switch to the
\textbf{\emph{global public}} \ac{MAC} address. The observed Handoff or Wi-Fi
Settings messages continue to use the properly randomized address. Nearby
messages are transmitted with global \ac{MAC} address continuously while these
other messages are being sent, then revert back to the randomized address. 

In practice, this means active use of Safari or Google Chrome on a
Handoff-enabled device will generate a constant flow of Nearby messages with
the global \ac{MAC} address.  Furthermore, the data fields of
both normal and these Handoff-concurrent Nearby messages match, allowing for
trivial correlation of the current randomized address to the device's real
\ac{MAC} address.  This passive observation technique entirely circumvents
\ac{BLE} \ac{MAC} randomization for any macOS device so long as the device has
another device associated to its iCloud account, Handoff has not been disabled,
and any Handoff-enabled application is in use. 

This macOS implementation flaw correlates the current random \ac{MAC} address,
public \ac{MAC} address, and current Handoff sequence number, all of which are
useful for tracking a device.  Further, knowledge of the public Bluetooth
\ac{MAC} provides an adversary with the ability to detect its presence in any
setting by initiating a connection attempt without needing sequence numbers at
all.  Finally, the Bluetooth \ac{MAC} is often offset $\pm 1$ from the Wi-Fi
\ac{MAC} address~\cite{ourFirstPaper, cache2007hacking}, enabling a passive
adversary to obtain an 802.11 identifier through \ac{BLE} collection. 

\vspace{-6mm}
\subsection{Device Stimulation \& Active Analysis}
\label{sec:active}
\vspace{-3mm}

A feature unique to \ac{BLE} is \ac{GATT}~\cite{gattover}, a framework to
discover, read, and write information to and from \ac{BLE} devices. Each
\ac{BLE} device that supports \ac{GATT} has a \ac{GATT} \emph{profile};
\ac{GATT} profiles define the type of \emph{services} that a device provides.
Within each service are well-defined \emph{characteristics}, and each
characteristic may contain several fields and values that describe that
characteristic. For example, ``Blood Pressure'' is a \ac{GATT} service that
describes blood pressure and other data related to blood pressure readings from
a monitor~\cite{gattspec}; within the blood pressure service are the ``blood
pressure measurement'' and ``intermediate cuff'' characteristics that describe
a blood pressure reading, and cuff pressure value during a blood pressure
reading, respectively. Each of these characteristics has a number of fields
with associated values, such as ``timestamp'' and ``heart rate'' that provide
further information to a remote device querying them. Services and
characteristics are uniquely identified by \acp{UUID}, in order to standardize
their meaning across a myriad of manufacturers.

\vspace{-1mm}
\centerline{\textbf{Device Fingerprinting}} 
While not a vulnerability in its own right, we discovered that Apple
devices support \ac{GATT} queries and provide detailed model information when
it is requested.  All Apple products we tested (iPhone, iPad, Apple Watch, and
MacBook) supported the ``Device Information'' service (\ac{UUID} \texttt{0x180A}),
and responded to its ``Model Number String'' characteristic (\ac{UUID}
\texttt{0x2A24}) with Apple's identifier string that uniquely identifies the
device model~\cite{iphonewiki,mbpmodels}. For example, an iPhone 7 we tested
returned the identifier ``iPhone9,1'', and a mid-2015 15-inch MacBook Pro with
Retina display returned ``MacBookPro11,4.'' We also note that while an
adversary must actively transmit data in order to retrieve the ``Model Number
String'' GATT characteristic, she may do so using a random source MAC address,
and the user of the device is unaware that they have received a GATT query, as
no prompt appears asking them to approve the data transmission.  Because of
this, it is exceptionally difficult to detect and prevent an adversary from
querying an Apple device model without disabling Bluetooth entirely. 
Although using the ``Device Information'' service to obtain the device model is
itself not novel (indeed, this is its purpose), in Section~\ref{sec:algo_id}, we
show that this knowledge assists in tracking individual devices. Devices that
respond with a different ``Model Number String'' than our tracking target can be
excluded from the pool of potential new identities after a change in random MAC
address.

The remainder of our active attacks focus on two complimentary types of Apple
Continuity messages -- Wi-Fi Settings and Instant Hotspot. Wi-Fi Settings
messages are generated when a user navigates to the Wi-Fi Settings page of
their iOS device, or when a user clicks on the Wi-Fi and network status icon on
the top of their screen on macOS. These messages trigger Instant Hotspot
messages in response from devices linked to the same iCloud account that can
act as a hotspot, and leads to the device appearing in the user's list of
available networks when viewed from a MacBook. 

\centerline{\textbf{Device Attributes}} 

In order to enumerate the possible values and meaning for the cellular field in
the Instant Hotspot message described in Section~\ref{sec:analysis}, we spoof
previously-captured Instant Hotspot messages from laboratory iPhones in
response to Wi-Fi Settings messages from a MacBook Pro on the same iCloud
account. By enumerating the possible values for the 1-byte cellular service
field and observing the type of service displayed for our spoofed device in the
laptop's available networks list, we were able to exhaustively classify each
value without having a device actually receiving that type of service
(\emph{e.g.}, LTE, 3G, EV-DO, \etc) We note that the source MAC address in our spoofed
messages must be a resolvable private random address that the MacBook can
resolve, for this reason, we choose to replay the source MAC observed.

\centerline{\textbf{Tracking}} 

We demonstrate that spoofing Wi-Fi Settings messages provides a
tangible benefit for an attacker. Because a device on the same iCloud
account that can provide Instant Hotspot service will respond without user
intervention when it receives a correct Wi-Fi Settings message from a laptop, we
are able to replay previously-captured Wi-Fi Settings messages hours later. As with
Instant Hotspot message spoofing, the source MAC address must be one that can be
resolved by the iPhone or iPad that can provide the hotspot service, and we note
that the iCloud ID field in the message changes daily at a fixed time per iCloud
account.  As such, these messages may be replayed for a maximum of 24 hours. 

\centerline{\textbf{Tracking}} 

Finally, we measure the effect of incoming SMS messages and phone calls on devices
running in order to determine whether an adversary with
knowledge of their target's phone number could stimulate their device in order
to discover whether it is in close proximity.

All of the iOS versions we tested began sending Handoff messages when a phone
call was accepted, or the Messages application was opened in response to an SMS
message. iOS 9 did not send Nearby messages, as Nearby messages are not a
feature of any iOS earlier than 10. In addition to sending Handoff messages when
a user takes an incoming call or opens the Messages app in response to an SMS
message, iOS 10 and 11 also send Active User Nearby messages in addition to the
Handoff messages that were sent in iOS 9. 

Concerningly, the latest iOS version, iOS 12, proved the most useful for
targeted tracking of users because it required \emph{no interaction on behalf
of the user} in order to determine whether a phone call was incoming or a text
had been received.  Because iOS 12 devices transmit Nearby messages constantly,
an attacker merely needs to observe a change in the Nearby message Action Codes.
Pre-iOS 12.3, a change from the locked state (Nearby
Action Code 3) to the transition state (Nearby Action Code 7), indicated an
incoming call or text, as the screen is
illuminated when a call or text is received. An adversary with the ability to
send a text message or initiate a call to a device can trivially identify the
device if it is in close proximity, defeating the private, random \ac{MAC}
address and allowing for the type of user tracking we outline in
Section~\ref{sec:algo_id}. Finally, the version of iOS at time of publication,
12.3, makes detecting an incoming phone or Facetime call even more trivial by
introducing Nearby Action Code 14; a device transmitting this Action Code is
in an active call. 

\vspace{-8mm}
\section{Evaluation}
\label{sec:algo_id}
\vspace{-3mm}

Next we evaluate how effectively the above flaws, in concert, can be
used by an adversary to track a device.

\vspace{-6mm}
\subsection{User Tracking}
\vspace{-3mm}

As described above, BLE devices periodically rotate their MAC address in order
to prevent tracking.  Ideally, the \ac{MAC} address is regularly set to a fresh
random value and potential adversaries eavesdropping on a device at different
times and/or locations will not be able to correlate these observations and
confidently identify them as belonging to the same device.  In our tests we
find that Apple devices rotate the addresses every 15 minutes, as recommended
by the BLE specification~\cite{corespec}.  However, the data leakage described
above can be used, in many cases, to defeat randomization and track a device
even through \ac{MAC} address changes.  

The main flaw which allows this is the fact that Handoff messages are sent out
regularly and with monotonically increasing sequence numbers.  The first
implication of this is if an adversary is present at the time that a \ac{MAC}
address is changed to a new random value, this transition is identifiable
because the sequence number and data fields of Handoff messages do not change
even when the \ac{MAC} address does.  We have observed that after a MAC address
change the sequence number stays constant until a new Handoff-related action is
performed by the user before continuing to increment.  

Similarly, the four-byte data field in iOS 11 and 12 Nearby messages, rather
than immediately rotating when the \ac{MAC} address changes, remains constant
for one to two frames after the \ac{MAC} address change.  As annotated in
Section~\ref{sec:nearby}, Nearby messages never stop transmitting in iOS 12 or
macOS Mojave, making this information leak a more powerful tracking method. 

\subsubsection{Longer Time Frames}
\vspace{-5mm}
Over longer time scales, it is very likely that an adversary will not be
present to observe every \ac{MAC} address change, and will eventually lose
track of the \ac{MAC} address of a particular device.  Due to the slow
monotonic increase of the sequence number and the relatively large sequence
number space it is possible to track devices by using the sequence number
itself as an identifier.  

\begin{figure*}[!htb]
\centering
  \includegraphics[width=.9\textwidth]{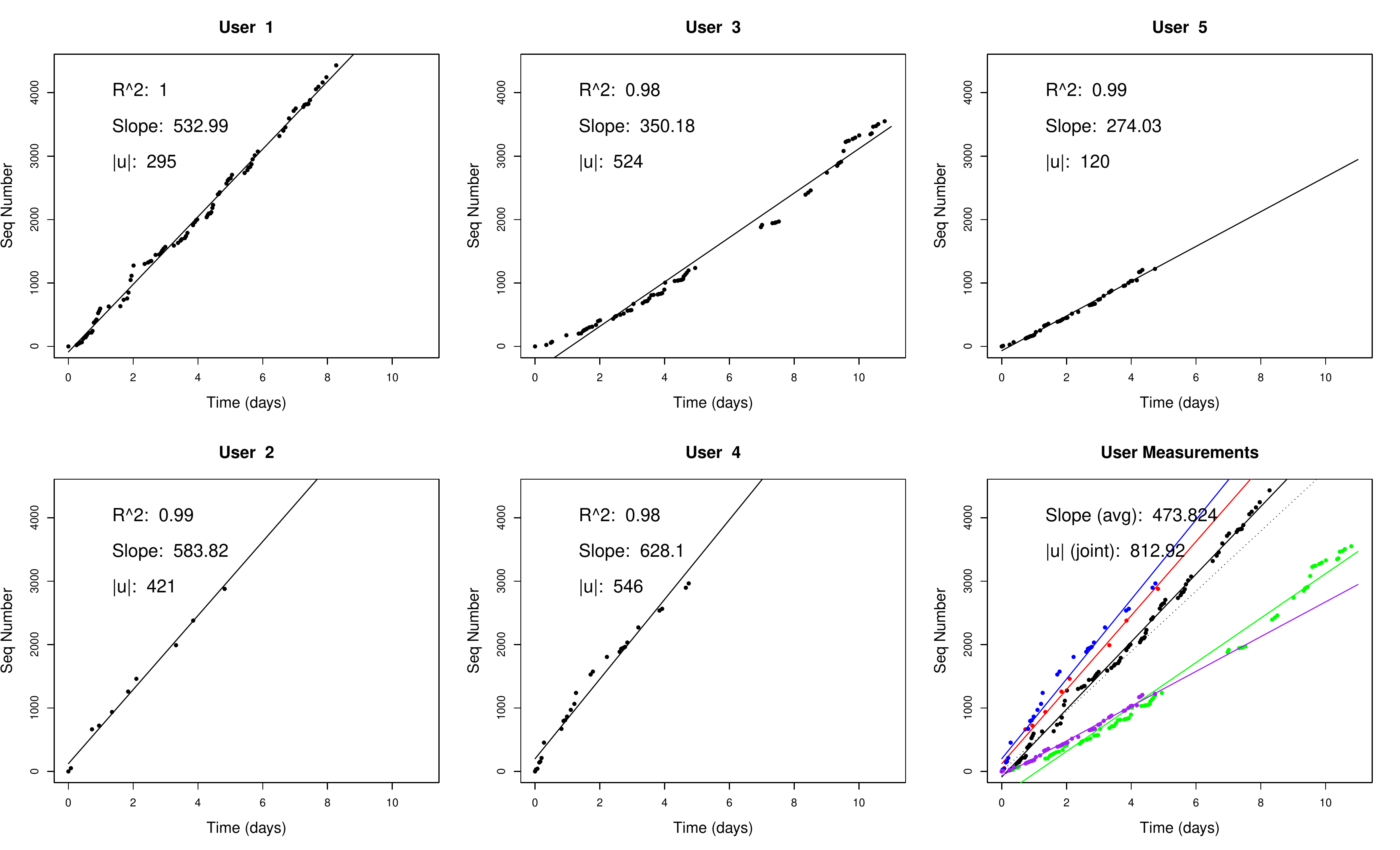}
  \vspace{-4mm}    
  \caption{Regressions of our collected data on sequence numbers. |u| is
calculated with 90\% prediction intervals}
  \label{fig:tracking-seqtrajectories}
  \vspace{-4mm}
\end{figure*}

To illustrate and quantify the effectiveness of this tracking capability we
performed four sets of measurements that demonstrate the predictability of
sequence numbers and the likelihood that a device can be uniquely identified in
a public place by their sequence number.  First, we measured how the sequence
numbers for devices owned by members of our research team increase over time
during regular daily use. Second, we measured the distribution of sequence
numbers in public places. Third, we estimated the probability of a second
device coincidentally having a sequence number close to the target,
representing a false positive for a tracking adversary.  Finally, we moved a
device belonging to a member of our research team through a variety of public
places over time and capture \ac{BLE} signals, attempting to identify this
targeted user's device.

\textbf{Sequence Number Trajectories}. Since sequence numbers increment when
specific user actions are taken on the device (use of a Handoff-enabled app,
device setting manipulation, SMS/call activity) we hypothesized the rate at
which sequence numbers increase is stable and predictable based on the usage
patterns of individual users. This would mean the sequence numbers of devices
may be predicted with high accuracy even when devices are not under
observation, and devices that leave an adversary's collection region may be
re-identified when returning to collection proximity. 

Five members of our research team (four students, one faculty)
used their iPhones normally over a period of one week including the weekend and
recorded their sequence numbers as close to hourly as possible. We present our
results in Figure~\ref{fig:tracking-seqtrajectories}, displaying the
trajectories as well as $|u|$, the size of the projected window of sequence
numbers for that user over time, as described below for accuracy estimates. We
expect sequence numbers to increase more quickly with the use of
Handoff-enabled apps on the device. In addition, we left a device on but unused
for the duration of the experiment and observed that its sequence number did
not increment. 

Given the predictable per-user slope of the observed sequence numbers, we
believe that over a period of a few days to a week the sequence number of a
particular device can be predicted with a high degree of accuracy.  We note
that while we have evidence to suggest this predictability occurs in certain
cases (among members of our research team in a given week), we only conjecture
that this is the case more generally. 

\textbf{Sequence Number Distributions}. Additionally, we took passive captures
of sequence numbers in the wild at four distinct public locations. We
hypothesize that sequence numbers are uniformly distributed in the space $[0,
65535]$ since, while sequence numbers begin at zero, they increase consistently
through regular use.  We passively collected \ac{BLE} signals and removed
redundant measurements by ignoring multiple Handoff messages from the same
\ac{MAC} address.  Our results are presented in Appendix~\ref{appendix}.

\textbf{Estimates}. In order to provide estimates on the accuracy of this tracking
method we use our measurement data and data from external sources wherever
possible. In cases where the data is unavailable we make conservative
independence assumptions, noting that in general more precise measurements of
the distributions required for these estimations will result in more accurate
tracking capabilities than we outline here.  

We estimate collision probabilities under our different attack models against
iPhones.  Our passive reconnaissance outlined in Section~\ref{sec:passive}
allow an adversary to bin devices by \ac{OS} version, while the active attacks
from Section~\ref{sec:active} allow the adversary to determine the granular
hardware submodel of a device as well. 

We use statistics from Mixpanel~\cite{mixpanelstats} to estimate the proportion
of Apple devices of each type in the wild as of February 25, 2019 and
statistics from Apple~\cite{iosStats} as of February 24, 2019 to estimate the
proportion of devices with each iOS installed.  We note that iPhone 7 is the
most common hardware model and in order to estimate the proportional size of
the largest bin we assume:

\begin{figure*}[!htb]
  \centering
  \includegraphics[width=.83\textwidth]{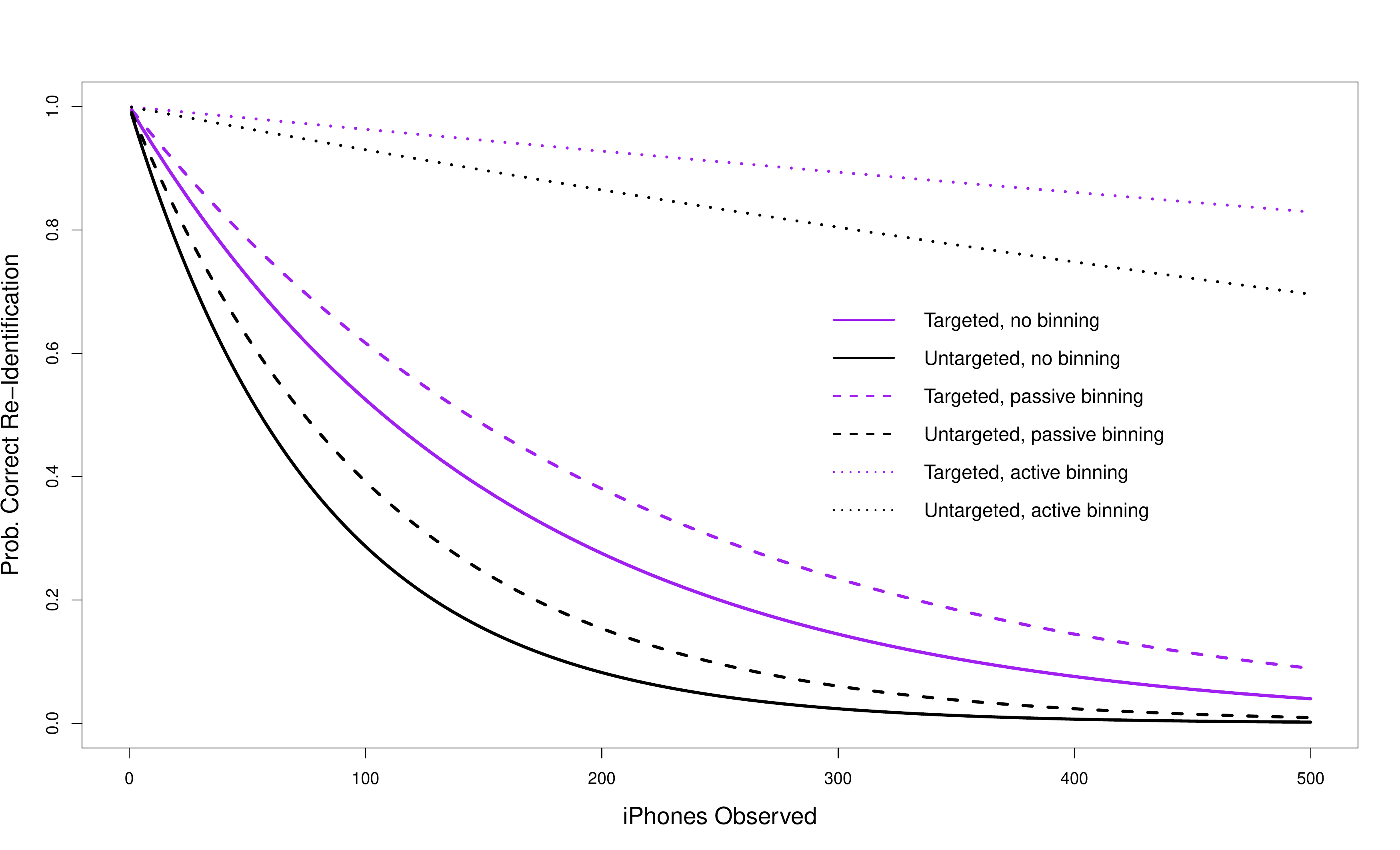}
  \vspace{-5mm}
  \caption{Re-identification accuracy under different attack scenarios}
  \label{fig:collisionprobs}
  \vspace{-5mm}
\end{figure*}

\begin{itemize}
  \item That the distributions of iOS version number and iPhone hardware
model are independent. Given this, we select the most prevalent iOS version,
version 12. 
  \item That iPhone 7 devices are split evenly between the two hardware
sub-models (iPhone9,3 vs iPhone 9,1)~\cite{iphonewiki} 
\end{itemize}

\vspace{-4mm}
For an estimate of the adversary's accuracy in re-identifying a device, we
consider two cases: targeted, where an adversary takes measurements of a
specific device and wishes to track it over time, and untargeted, where an
adversary does not have data but attempts to determine if a device is a
previously observed device or a new one. The adversary must calculate a window
of plausible sequence numbers associated with a previously observed device to
determine whether a new measurement is the previous device or a new one, and
the adversary may incorporate knowledge about the expected use patterns of its
target to determine this window. In the targeted case, where an adversary has
made many measurements of a specific user over time we estimate as the size of
the window $u = 421$ the median size of the largest $90\%$-prediction interval
size from our five experiments, and in the untargeted case, where an adversary
assumes the device increment rate sits between our largest and smallest
observed rates we take the more conservative convex closure of all $90\%$
prediction intervals, $[u_{\textsc{min}},u_{\textsc{max}}]$ which gives $u =
813$. 

In this setting we can calculate the probability that a target device will be
the only one in a given location with a sequence number in the target window as
\[ \left(1 - \frac{u}{65536}\right)^n \] where $n$ is the number of devices
that are identically configured to the target (i.e., look the same according to
all of our binning techniques).  

We calculate the likelihood that a user is correctly re-identified (i.e., that
an in-bin sequence number collision does not occur) under each possible attack
scenario in Figure~\ref{fig:collisionprobs}.

\begin{figure*}[!htb]
  \centering
  \includegraphics[width=.9\textwidth]{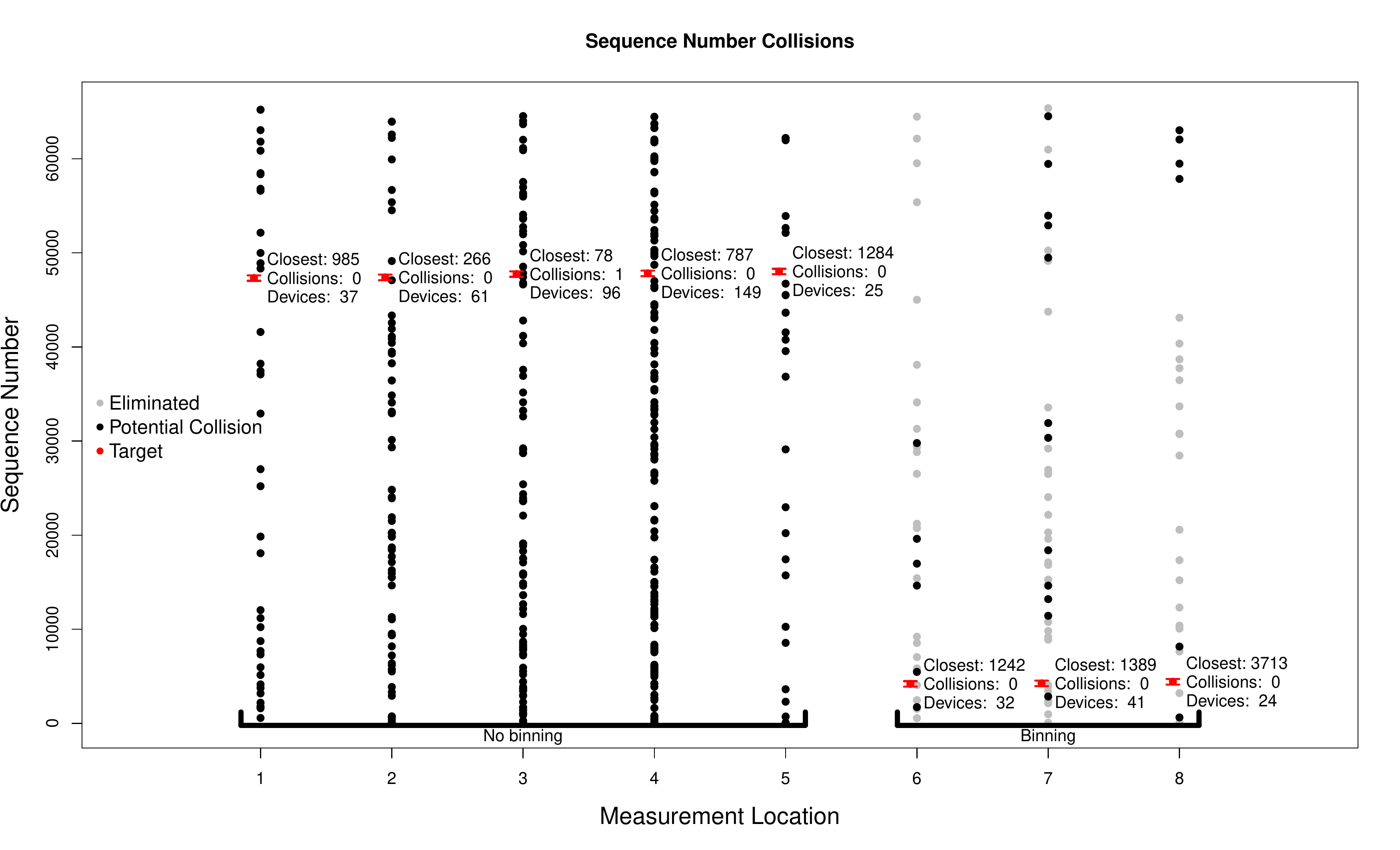}
  \vspace{-4mm}
  \caption{Collision Measurements}
  \label{fig:collision-measurements}
  \vspace{-5mm}
\end{figure*}

\textbf{Sequence Number Collisions}.  We performed eight short measurements in
public locations, presented in Figure \ref{fig:collision-measurements}, to
verify the viability of our tracking methods in a real-world scenario. Each
measurement was in a different location in the same city, lasting up to 40
minutes each. User 1 from our trajectory experiments took their iPhone 6S with
a known sequence number and moved to each location and another researcher
captured nearby Handoff traffic, with a third applying methods described in
Section~\ref{sec:active} in an attempt to identify the hardware models of
devices in close proximity. In the last three experiments, a different user was
targeted who had an iPhone 6. Nearly every device we observed in the wild was
running iOS 12 so we do not include software binning in our results. We note
that our experimental setup was not able to accurately identify hardware models
for most devices in the first five experiments, so we ignore hardware profiles
for these measurements and consider them a test of tracking without binning.
For measurements six, seven, and eight we made significant improvements to our
collection software and identified the hardware model of $81\%$, $70\%$, and
$75\%$ of devices respectively.  The devices with hardware profiles that are
unknown or that match the target device we take as collisions, and we note that
in total 90 of 465 total devices were successfully binned by hardware, and of
these 90, four were in the same bin as the target device. 

Unlike our estimates, these measurements include devices of all types (watches,
notebooks, etc) while in the estimates above our unit of measure is observed
number of iPhones. These experiments test the targeted setting, so we use as
our window the median size from our trajectory experiments $|u| = 421$. 

We conclude, from our estimates and measurements:
\bi
  \item Active adversaries can reliably re-identify even the most common
devices over time in public places without any long-term
targeted data collection. 
  \item Passive adversaries can reliably re-identify devices that are less
common, by targeting specific users over time, or when the number of observed
devices is low.
  \item Negative results about the presence of a given device are highly
accurate, even for passive adversaries. 
\ei

\begin{comment}
\vspace{-4mm}
\begin{table}[!h]
  \center
  \caption{Distribution of iPhones in the Wild}
  \label{table:iphones-table}
  \begin{tabular}{l|r}
    \toprule
      Device & \%  \\ 
    \midrule
	iPhone XR & $4.65$  \\
	iPhone XS & $3.59$  \\
	iPhone XS Max & $4.34$ \\
	iPhone X & $11.85$  \\
	iPhone 8 & $10.47$  \\
	iPhone 8 Plus & $10.49$ \\
	iPhone 7 & $15.49$  \\
	iPhone 7 Plus & $9.33$ \\
	iPhone SE & $3.87$ \\
	iPhone 6s Plus & $3.43$ \\
	iPhone 6s & $10.24$ \\
        All other iPhone models & $12.28$ \\ 
    \bottomrule
  \end{tabular}
\end{table}
\vspace{-4mm}
\end{comment}

\subsection{Behavioral Data}
\vspace{-4mm}

We note the sequence number is a measurement of user interactions with the
device that is persistent over time and regularly broadcast over \ac{BLE}.
Similarly, Nearby Action Codes explicitly broadcast information on how a device
is being used in the moment. This means \ac{BLE} traffic from Apple devices
provides multiple sources of private behavioral data. In particular, this
allows adversaries to make statistical inferences about the quantity of Handoff
device interactions that have occurred between measurements. Behavioral
statistics of this nature present an area for interesting future work, though
we recommend mitigation strategies to Apple for future software versions
changes so that this information can no longer be collected. 

\vspace{-8mm}
\section{Remediation}
\vspace{-3mm}

In this section, we recap the flaws we discovered and discuss how
they could be addressed.  In most cases, there is nothing that a user can do to
protect themselves.  The fixes have to be addressed by Apple at the kernel or
firmware level.  We believe the most straightforward solution to most of these
flaws is to either remove the plaintext information that is being leaked, if it
is not necessary, or encrypt it with the shared encryption key used across all
devices on the same iCloud account.  We have notified Apple of our findings and
hope to work with them to address the issues we have identified.

\vspace{-8mm}
\paragraph*{macOS Global MAC Address} 
We believe that this behavior is an unintended bug resulting in a serious
tracking vulnerability. As such it should be easily correctable and we
recommend Apple release an updated patch correcting the behavior.

\vspace{-8mm}
\paragraph*{MAC randomization}
In general, the frequency with which MAC addresses are rotated (every 15
minutes) is a substantial flaw.  If an adversary is within range of the device
when it rotates, even fixing all of the flaws we have discussed so far, it is
not usually difficult to link two subsequent randomized MAC addresses.  This is
because the rotation event is so infrequent that an adversary can observe that 
one MAC address stops transmitting at precisely the same time that another one
starts and deduce that they are actually the same device.  Even worse, since it
always happens every 15 minutes, it can be predicted like clockwork.  The ideal
solution would be for every frame to have a new randomized MAC, but if that is
too burdensome then devices should at least have a shorter period of rotation,
and that rotation should be done stochastically instead of on a fixed timer.

\vspace{-8mm}
\paragraph*{GATT commands}  
An adversary can use GATT to query \ac{BLE}-enabled Apple devices for their
exact model even when their MAC address is being randomized.  To reduce
identifiability of devices, iOS could respond to this query with a less
specific identifier (i.e., just ``Apple'' or ``iOS''), or not respond to it at
all since it is optional.  

\paragraph*{Nearby messages}
The Nearby messages are responsible for the most consistent behavioral leakage.
Since there is an Action Code for when the device locks, when the device
unlocks, when the user interacts with the device and another one that 
constantly broadcasts while the screen is on, a passive adversary can always
tell whether the device is:\\

\vspace{-4mm}
\bi
  \item Locked
  \item Unlocked and being actively used
  \item Unlocked but not being actively used 
  \item Currently in a phone or Facetime call
\ei
\vspace{-4mm}

Encrypting the Action Codes would effectively stop someone from learning this
behavior since the Nearby messages (at least in iOS 12 and macOS Mojave) are
sent constantly at a regular interval.  If the Code is hidden then all that
could be seen is a regular transmission of a message with no link to specific
behavior.

\vspace{-8mm}
\paragraph*{Handoff messages}
The main vulnerability we have discovered in Handoff messages is the presence
of a monotonically increasing two-byte sequence number which allows an
adversary to defeat MAC address randomization and track a device over a long
period of time.  However, there appears to be no need for a sequence number in
these frames at all.  Handoff messages are sent over the \ac{BLE} advertising
channel, meaning that these messages are not part of an established
communication link between devices.  If a message is missed, there is no
mechanism for a receiving device to ask for it to be repeated anyway.  The sequence
numbers should be removed.

We also found that the clipboard status was leaked in the clear.  This
message should be encrypted.  Additionally, the data field in each Handoff
message (which we believe \emph{is} encrypted) stays constant when a device's
MAC address rotates, allowing an adversary to link two subsequent MAC
addresses.  Therefore, this encryption should be refreshed whenever the MAC
address changes.

\vspace{-8mm}
\paragraph*{Wi-Fi Settings and Instant Hotspot}
When a user opens the Wi-Fi settings page on an idevice, a Wi-Fi Settings frame
is sent.  Nearby devices that are capable of Instant Hotspot respond with an
Instant Hotspot frame containng information such as battery life of the
device, how many bars of cell service it has, and what type of cellular
network it is connected to.  The Wi-Fi Settings frame also contains a cleartext
iCloud ID, shared amongst all devices on the same iCloud account.  This
not only leaks information about the state of the phone, but may also allow an
adversary to link different devices to the same owner.

The information about the device (battery life, etc.) should be encrypted.
Similarly, there is no reason to use a static identifier that facilitates
tracking and linking of devices.  The identifier could also be encrypted with
fresh randomness so that it does not appear the same in two different Wi-Fi
Settings frames.

iOS devices also transmit a message when joining a network which includes a
hash of the SSID that they are joining.  Again, this hash should incorporate
the shared encryption key so that it is cannot be easily reversed by an
adversary.

\vspace{-8mm}
\section{Conclusion}
\label{sec:conclusion}
\vspace{-3mm}

In this work we present several flaws in Apple's Continuity protocols which
leak device and behavioral data to nearby listeners.  Individually, each flaw
leaks a small amount of information, but in aggregate they can be used to
identify and track devices over long periods of time, despite significant
efforts in other parts of the \ac{BLE} protocol to prevent this scenario (MAC
randomization). Device designers use short range wireless communications
protocols like \ac{BLE} to generate a fluid user experience across many 
devices, a very attractive product feature given the proliferation of
connected mobile devices.  However, securely deploying these technologies presents
difficult privacy challenges. Not only do these short-range transmissions
inherently leak location information, it seems the most practical uses for 
these communications are real-time activity updates which are themselves 
user data. Finally, the short range nature of these
technologies means that signals on two independent channels can often be 
identified as coming from the same device, meaning that privacy
vulnerabilities in one wireless domain could entirely trivialize 
well-implemented safeguards in another as we have presented here. 

\vspace{-8mm}
\section*{Acknowledgment}
\vspace{-4mm}
We would like to thank the Apple privacy team who provided prompt
feedback and guidance.  Views and conclusions are those of the authors and
should not be interpreted as representing the official policies or position of
the U.S. Government.  The author’s affiliation with The MITRE Corporation is
provided for identification purposes only, and is not intended to convey or
imply MITRE’s concurrence with, or support for, the positions, opinions or
viewpoints expressed by the author. The authors additionally thank Dane Brown,
Caroline Sears, Peter Ryan, and Robert Beverly for technical assistance and
feedback.

%\newpage

%\vspace{2mm}
\small{
 \bibliography{refs}
 \bibliographystyle{abbrvnat}

}

\appendix
\section{Histograms of sequence numbers captured in public locations}
%JM:reword title
\label{appendix}

\begin{minipage}{\textwidth}
%\begin{figure*}[h!]
	\centering
	  \includegraphics[width=.98\textwidth]{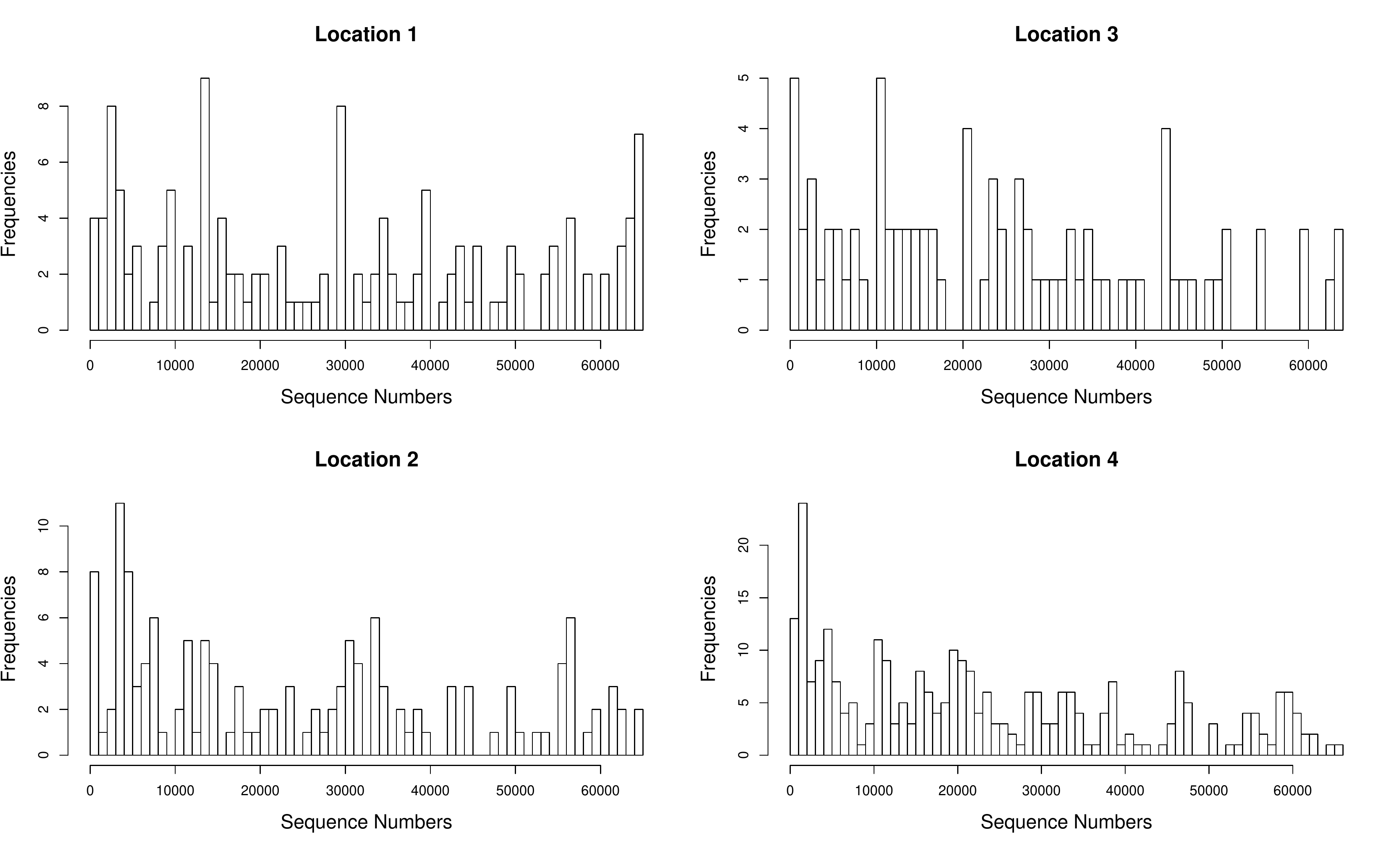}
	  \vspace{-2mm}
	  %\caption{Histograms of sequence numbers captured in public locations}
	  %\label{fig:seqnum-wildcaptures}
%\end{figure*}
\end{minipage}

\begin{acronym}
  \acro{AP}{Access Point}
  \acro{AWDL}{Apple Wireless Direct Link}
  \acro{BR}{Basic Rate}
  \acro{BLE}{Bluetooth Low Energy}
  \acro{CAC}{Common Access Card}
  \acro{CCA}{Client Certificate Authentication}
  \acro{DoD}{Department of Defense}
  \acro{EC}{Elliptic Curve}
  \acro{EDR}{Enhanced Data Rate}
  \acro{GATT}{Generic Attributes}
  \acro{GB}{gigabyte}
  \acro{IE}{Information Element}
  \acro{IEEE}{Institute of Electrical and Electronics Engineers}
  \acro{IMEI}{International Mobile Equipment Identities}
  \acro{IMSI}{International Mobile Subscriber Identities}
  \acro{IRK}{Identity Resolving Key}
  \acro{IRB}{Institutional Review Board}
  \acro{MAC}{Media Access Control}
  \acro{MiTM}{Man in The Middle}
  \acro{OS}{Operating System}
  \acro{PII}{Personally Identifiable Information}
  \acro{PKI}{Public Key Infrastructure}
  \acro{RSSI}{Received Signal Strength Indicator}
  \acro{SSID}{Service Set IDentifier}
  \acro{TLS}{Transport Layer Security}
  \acro{USNA}{U.S. Naval Academy}
  \acro{UUID}{Universally Unique IDentifier}
\end{acronym}

\end{document}